\documentstyle[preprint,aps,eqsecnum,tighten,epsf]{revtex}

\begin{document}
\preprint{\vbox{\baselineskip=12pt
\rightline{SNUTP 96-117}}}

\draft
\title{4-Dimensional Kaluza-Klein
Approach to General \\ 
Relativity in the (2,2)-Splitting of Spacetimes}
\author{ J.H. Yoon\thanks{Electronic address: 
yoonjh@cosmic.konkuk.ac.kr}}
\address{Department of Physics\\
Kon-Kuk University, Seoul 143-701, Korea}

\maketitle
\thispagestyle{empty}

\begin{abstract}
We propose a Kaluza-Klein approach to general relativity 
of 4-dimensional spacetimes. This approach is based 
on the (2,2)-splitting of a generic 4-dimensional spacetime, 
which is viewed as a local product of 
a (1+1)-dimensional base manifold and a 2-dimensional fibre 
space. In this Kaluza-Klein formalism we find that
general relativity of 4-dimensional spacetimes 
can be interpreted in a natural way as a (1+1)-dimensional 
gauge theory. In a gauge where the (1+1)-dimensional metric 
can be written as the Polyakov metric, the action is describable 
as a (1+1)-dimensional Yang-Mills type gauge theory action
coupled to a (1+1)-dimensional non-linear sigma field and 
a scalar field, 
subject to the constraint
equations associated with the diffeomorphism invariance of the
(1+1)-dimensional base manifold. Diffeomorphisms 
along the fibre directions show up as the Yang-Mills type
gauge symmetries, giving rise to the Gauss-law constraints.  
We also present the Einstein's equations in the Polyakov
gauge and discuss them from the (1+1)-dimensional gauge 
theory point of view. 
Finally we show that this Kaluza-Klein formalism 
is closely related to the 
null hypersurface formalism of general relativity. 
\end{abstract}


\begin{section}{Introduction}
\label{s1}

The exact correspondence of the Euclidean self-dual Einstein's 
equations to the equations of motion of 2-dimensional non-linear 
sigma models with the target space as the area-preserving 
diffeomorphism of 2-surface\cite{park} has inspired us 
to look further into the intriguing question whether 
the full-fledged general relativity of 4-dimensional 
spacetimes can be also formulated as a certain 
(1+1)-dimensional field theory. Recently we have shown that 
such a description is indeed possible, by rewriting
the Einstein-Hilbert action of general relativity of 
4-dimensional spacetimes as a 4-dimensional Kaluza-Klein (KK) 
action\cite{yoon,yoona,yoonb}
in the framework of the (2,2)-decomposition of spacetimes. 
In this KK approach, the 4-dimensional 
spacetime is viewed, at least for a finite range of spacetime,
as a locally fibred manifold that consists of a 
(1+1)-dimensional base manifold and a 2-dimensional 
``auxiliary" fibre space.
Remarkably, it turns out that the KK approach, 
which is often thought a means of dimensional reduction 
to 4-dimensional spacetimes from higher dimensions
via isometries,
can be made a useful tool for studying 
general relativity of a generic spacetime of 4-dimensions
from the perspective of lower dimensional 
spacetimes such as the (1+1)-dimensional spacetime, for instance.

There are certain advantages of this 4-dimensional KK approach 
to general relativity in the (2,2)-splitting, which led us to
develop this formalism. Here we list a few of them. 
First of all, 
in (1+1)-dimensions, there exist a number of field 
theoretic methods recently developed thanks to the 
string-related theories. Hopefully the rich mathematical methods
suitable for (1+1)-dimensional field theories 
might also prove useful in studying certain aspects of 
general relativity of 4-dimensional spacetimes, 
provided that a sensible (1+1)-dimensional field theory description 
of such spacetimes is available.
In fact, in this KK approach based on the (2,2)-splitting of 
spacetime, it will be seen that general relativity can be
understood in a natural way as a (1+1)-dimensional Yang-Mills 
type gauge theory coupled to a (1+1)-dimensional non-linear 
sigma field and a scalar field, 
with built-in symmetries and interactions prescribed by the
diffeomorphism invariance of the associated 4-dimensional
spacetimes. 

Second, this Yang-Mills type gauge theory description 
of general relativity simplifies considerably 
certain issues concerned with the constraints.
In Yang-Mills type theories, it is well-known that 
the Gauss-law constraints associated with the Yang-Mills 
gauge invariance can be made trivial, 
if we consider physical, gauge invariant quantities only.
Thus, in principle, one might expect that
the problem of solving constraints of general relativity 
could be made {\it trivial}, 
at least for some of them, if such a gauge theory description 
is possible. In the (2,2)-KK approach, 
the relevant constraints that can be treated this way
are those associated with the group of diffeomorphisms of 
the 2-dimensional fibre space, 
{\it not} the entire 4-dimensional spacetime.
Of course the remaining two constraints must be 
studied separately\cite{solo}.

Third, in the null hypersurface formalism, it is well-known 
that the exact gravitational wave
degrees of freedom reside in the conformal 2-geometry of 
the wave front 2-surface, transverse to the propagation 
direction of the gravitational wave.
This identification of the exact gravitational degrees of freedom
has been particularly useful in the analysis of 
gravitational waves in the asymptotically flat 
spacetimes\cite{bondi,sach}, and
in the asymptotic quantization\cite{asht,ashta,carlos} and
in the canonical analysis of the null hypersurface 
formalism\cite{helmut,gold,torr}, in the hope of
getting quantum theory of gravity
by quantizing the true physical
degrees of freedom of gravitational field.
It turns out, as we shall see later, 
that the (2,2)-KK formalism is in essence  
a {\it novel} reinterpretation of the null hypersurface 
formalism of 4-dimensional vacuum general relativity 
from a (1+1)-dimensional point of view, 
using the Kaluza-Klein ideas. Therefore we expect that 
this (2,2)-KK formalism will be most 
appropriate for studying problems involving gravitational waves 
such as scatterings of gravitational waves, 
and quantizing gravitational waves, or 
perhaps for quantum gravity.

Fourth, it is well-known that the null hypersurface formalism, 
particularly the Newman-Penrose formalism, is outstanding
as a method of finding exact solutions of the Einstein's equations.        
We expect that this feature will also survive in the 
(2,2)-KK formalism,
in view of the manifest correspondence of both formalisms, as is shown 
in section \ref{s5} and \ref{s6}. 

Last, but not the least, this (2,2)-KK formalism brings about
{\it unusual} conceptual issues related to 
the interpretation of general relativity.
Namely, this (1+1)-dimensional description of 4-dimensional spacetimes
allows us to {\it forget} about the 4-dimensional spacetime 
picture of general relativity:
instead, it enables us to study the subject much 
the way as we study Yang-Mills type gauge theories 
coupled to matter fields in (1+1)-dimensions, where
the Yang-Mills gauge symmetry is an infinite dimensional 
group of diffeomorphisms of the 2-dimensional fibre space. 
Apart from the technical questions 
such as how to deal with such an infinite dimensional group of 
diffeomorphisms, this (1+1)-dimensional field theory description 
certainly puts 4-dimensional spacetime physics into 
a completely {\it new} perspective, from which a number of 
interesting speculations on general relativity of 4-dimensional 
spacetimes could be put forward.

In view of these advantages of the KK formalism 
in the (2,2)-splitting and the related 
null hypersurface formalism, it therefore seems worth 
exploring the (2,2)-KK formalism carefully
in order to see what could be learned more about general relativity
of 4-dimensional spacetimes.
This paper is organized as follows. 
In section \ref{s2}, we shall outline 
the general (2,2)-decomposition 
of 4-dimensional spacetimes, and introduce the KK variables
{\it without} assuming any spacetime Killing vector fields.
It will be seen that the Lie derivative, rather
than the metric-compatible covariant derivative, 
appears naturally in this formalism. 
We defer the details of computations to Appendix \ref{a2}. 

In section \ref{s3}, we shall introduce a coordinate condition, 
known as the Polyakov gauge in the literatures on 
(1+1)-dimensional field theories.
In the Polyakov gauge, the scalar curvature of 4-dimensional 
spacetime is found to assume a surprisingly simple
form that permits a natural (1+1)-dimensional field theoretic
interpretation.

In section \ref{s4}, we shall present the Einstein's equations
in the Polyakov gauge in terms of the KK variables. The natural 
time coordinate in this formalism is found to be
the {\it affine } parameter of the null geodesics. 
It is worth noting that one of the Einstein's
equations in this formalism can be written 
in a form of the Schr\"{o}dinger equation, 
reminiscent of the Brill wave equation.

In section \ref{s5}, a  review of the null hypersurface 
formalism is 
given, and the relation between the KK variables 
and the null tetrads is explicitly shown. 
Then, in section \ref{s6}, 
we shall establish the correspondence between the KK field 
equations and the Einstein's equations,
using the well-known classification of the Einstein's equations 
in the null hypersurface formalism. We summarize this paper
in section \ref{s7}.

In Appendix \ref{a1}, we shall find the transformation properties 
of the KK variables {\it without} assuming any spacetime Killing 
vector fields, and identify the KK scalar, 
tensor and gauge fields in the absence of such Killing vector fields.
In Appendix \ref{a2}, we shall present the prescription of computing 
the scalar curvature of 4-dimensional spacetimes, using the 
KK variables. The intermediate steps
of computations are also presented, in the hope that
computations can be relatively easily checked.
We shall also present the surface terms in the action
integral for future references, although we shall not use them
in this paper. In Appendix \ref{a3}, 
it is shown that $\kappa^{2}$, which appears in 
the discussion of the Einstein's equations in section \ref{s4}, 
is positive definite. 

\end{section}

\begin{section}{Kaluza-Klein Formalism in the (2,2)-Splitting of 
      Spacetimes}
\label{s2}

In this section we shall describe the general (2,2)-decomposition 
of 4-dimensional spacetimes 
from the KK perspective, where spacetime
is viewed as a {\it locally} fibred manifold, i.e. 
a local direct product of
a (1+1)-dimensional {\it base} manifold $M_{1+1}$ and 
a 2-dimensional {\it fibre} space $N_{2}$, 
for which we introduce two pairs of basis vector fields 
$\partial_{\mu}=\partial / \!  \partial x^{\mu}$ $(\mu=0,1)$ and  
$\partial_{a}=\partial / \!  \partial y^{a}$$ (a=2,3)$, 
respectively. Let us denote by $\hat{\partial}_{\mu}$
the vector fields tangent to the horizontal space, 
orthogonal to the fibre space $N_{2}$.
These horizontal vector fields can be written as linear combinations
of $\partial_{\mu}$ and $\partial_{a}$, 
\begin{equation}
\hat{\partial}_{\mu}=\partial_{\mu} 
- A_{\mu}^{\ a}\partial_{a},                \label{combi}
\end{equation}
where $A_{\mu}^{\ a}$ are the coefficient functions of the 
linear combinations.
Let us denote by  $\gamma_{\mu\nu}$ the metric 
on the horizontal space, and by $\phi_{a b}$ 
the metric on $N_{2}$,  respectively.
Then, without loss of generality, the line 
element of the 4-dimensional spacetime can be written 
as\cite{cho,mtw}
\begin{equation}
ds^{2}=\gamma_{\mu\nu}dx^{\mu}dx^{\nu} 
+ \phi_{a b}\Big( dy^{a} + A_{\mu}^{\ a}dx^{\mu} \Big)
            \Big( dy^{b} + A_{\nu}^{\ b}dx^{\nu}  \Big),\label{gen}
\end{equation}
or in components, 
\begin{eqnarray}
g_{AB}=\left(
\begin{array}{cc}
\gamma_{\mu\nu} +\phi_{a b}A_{\mu}^{\ a}A_{\nu}^{\ b} 
            & \phi_{a b}A_{\mu}^{\ b} \\
\phi_{a b}A_{\nu}^{\ a} & \phi_{a b}
\end{array}\right),            \label{gena} 
\end{eqnarray} 
where $A=(\mu, a)$.
Formally this metric looks similar to the
``dimensionally reduced" metric in standard KK theories,
but in fact it is quite different.
In the standard KK reduction we normally assume isometries, and
the dimensional reduction is made by projection
along the directions generated by these isometries. 
There the fields $A_{\mu}^{\ a}$ are identified 
as the KK gauge fields associated with the isometry group. 
In this paper, 
we do {\it not} assume such isometries: nevertheless, 
it turns out that the KK idea still works. To see this, 
let us consider the following coordinate transformations
of $N_{2}$ (diff$N_{2}$), while keeping $x^{\mu}$ constant, 
\begin{eqnarray}
& &y^{' a}=y^{' a} (x,y),    \nonumber\\
& &x^{' \mu}=x^{\mu},                      \label{new}
\end{eqnarray}
where $y^{' a} (x,y)$ is an arbitrary function of 
$x^{\mu}$ and $y^{a}$.
Under these transformations, we find the metric coefficients 
transform as (see Appendix \ref{a1}),
\begin{eqnarray}
& &\phi'_{a b}(x',y')={\partial y^{c} \over \partial y^{' a}}
   {\partial y^{d} \over \partial y^{' b}}
   \phi_{c d}(x,y),                          \label{phi}\\
& &A_{\mu}^{\ 'a}(x',y')={\partial y^{' a} \over \partial y^{c}}
   A_{\mu}^{\ c}(x,y) -\partial_{\mu} y^{' a}(x,y), \label{newtran}\\
& &\gamma '_{\mu\nu}(x',y')=\gamma _{\mu\nu}(x,y),   \label{gam}
\end{eqnarray}
which follows from the invariance of 
the line element (\ref{gen}) under the diff$N_{2}$ transformations. 
Under the corresponding infinitesimal transformation
\begin{eqnarray} 
& &\delta y^{a}=\xi^{a} (x,y) 
\hspace{1cm} (0(\xi^{2}) \ll 1),     \nonumber\\
& &\delta x^{\mu}=0,                   \label{var}
\end{eqnarray} 
where $\xi^{a}(x,y)$ is an arbitrary function of 
$x^{\mu}$ and $y^{a}$, the fields transform as
\begin{eqnarray}
\delta \phi_{a b}(x,y)&=&-\xi^{c}\partial_{c}\phi_{a b}
   -(\partial_{a}\xi^{c})\phi_{c b}
   -(\partial_{b}\xi^{c})\phi_{a c}    \nonumber\\
&=&-\pounds_{\xi}\phi_{a b}             \nonumber\\
&=&-[ \xi, \phi ]_{{\rm L} a b},       \label{var2}\\
\delta A_{\mu}^{\ a}(x,y)&=&-\partial_{\mu}\xi^{a}+A_{\mu}^{c}
    \partial_{c}\xi^{a} 
    -\xi^{c}\partial_{c}A_{\mu}^{\ a} \nonumber\\
&=&-\partial_{\mu}\xi^{a}+ \pounds_{A_{\mu}}\xi^{a}\nonumber\\ 
&=&-\partial_{\mu}\xi^{a} + [A_{\mu},\ \xi ]_{{\rm L}}^{a}, 
                   \label{var1}\\
\delta \gamma_{\mu\nu}(x,y)
&=&-\xi^{c}\partial_{c} \gamma_{\mu\nu} \nonumber\\     
&=&-\pounds_{\xi}\gamma_{\mu\nu}      \nonumber\\
&=&-[ \xi, \gamma_{\mu\nu}]_{{\rm L}}.      \label{var3}
\end{eqnarray}
Here $\xi=\xi^{a}(x,y)\partial_{a}$ and 
$A_{\mu}=A_{\mu}^{\ a}(x,y)\partial_{a}$
are vector fields tangent to $N_{2}$, 
and the brackets with the subscript ${}_{\rm L}$ 
stand for the {\it Lie} 
derivatives\footnote{Here the Lie derivative is defined such that
it acts on the fibre space indices $a, b, c, \cdots $ only.}.
Associated with this diff$N_{2}$ transformation, the 
diff$N_{2}$-covariant derivative $D_{\mu}$ is defined naturally as
\begin{equation}
D_{\mu}=\partial_{\mu}-[A_{\mu}, \ \ ]_{{\rm L}}.       \label{cd} 
\end{equation}
With this definition, it follows that
\begin{equation}
\delta A_{\mu}^{\ a}=-D_{\mu}\xi^{a},
\end{equation}
which clearly indicates that $A_{\mu}^{\ a}$ are
the gauge fields valued in 
the infinite dimensional Lie algebra of diff$N_{2}$. 
Moreover, a {\it covariant} derivative
of the field $\phi_{ab}$ is defined as
\begin{eqnarray}
D_{\mu}\phi_{a b}
&=&\partial_{\mu}\phi_{a b} 
- [A_{\mu}, \phi ]_{{\rm L} a b}       \nonumber\\
&=&\partial_{\mu}\phi_{a b}
-A_{\mu}^{\ c}\partial_{c}\phi_{a b}
-(\partial_{a}A_{\mu}^{\ c})\phi_{b c}
-(\partial_{b}A_{\mu}^{\ c})\phi_{a c}.
\end{eqnarray}
The field strength $F_{\mu\nu}^{\ \ a}$ corresponding 
to $A_{\mu}^{\ a}$ can now be defined as
\begin{eqnarray}
F_{\mu\nu}^{\ \ a}
&=&\partial_\mu A_{\nu} ^ { \ a}-\partial_\nu A_{\mu} ^ { \ a} 
- [A_{\mu}, A_{\nu}]_{\rm L}^{a}      \nonumber\\
&=&\partial_\mu A_{\nu} ^ { \ a}-\partial_\nu A_{\mu} ^ { \ a}
-A_{\mu}^{\ c}\partial_{c}A_{\nu}^{\ a}
+A_{\nu}^{\ c}\partial_{c}A_{\mu}^{\ a}.           \label{fiea}
\end{eqnarray}
As is shown in Appendix \ref{a1}, 
$D_{\mu}\phi_{a b}$ and $F_{\mu\nu}^{\ \ a}$
transform {\it covariantly} under the
infinitesimal transformation (\ref{var}),
\begin{eqnarray}
\delta ( D_{\mu}\phi_{a b} )
&=&-\xi^{c}\partial_{c}(D_{\mu}\phi_{a b})
   -(\partial_{a}\xi^{c})(D_{\mu}\phi_{b c})
   -(\partial_{b}\xi^{c})(D_{\mu}\phi_{a c}) \nonumber\\
&=&-[ \xi, D_{\mu}\phi]_{{\rm L} ab}, \label{fifc}\\
\delta F_{\mu\nu}^{\ \ a}&=&
-\xi^{b}\partial_{b}F_{\mu\nu}^{\ \ a} 
+F_{\mu\nu}^{\ \ b} \partial_{b} \xi^{a} \nonumber\\
&=&-[\xi, F_{\mu\nu}]_{\rm L}^{a}.         \label{fifa}
\end{eqnarray}
It must be marked here that, in the
(2,2)-KK formalism, 
the Lie derivative, rather than the covariant derivative,
appears naturally\footnote{
There are occasions when the Lie derivative
becomes the natural derivative operator. For instance,
at null infinity ${\cal I}^{+}$, the natural 
derivative operator is the Lie derivative, rather than the 
covariant derivative $\nabla_{A}$, because the metric 
on ${\cal I}^{+}$ is degenerate with signature 
$(0,+,+)$\cite{geroch}.}.
Particularly, in the defining equation (\ref{cd}) of 
the diff$N_{2}$-covariant derivative, the Lie derivative 
along the vector fields $A_{\mu}$ is formally identical to 
the Lie commutator through which the {\it minimal}
couplings to gauge fields are introduced in conventional 
gauge theories. Together with the fact that 
the Lie derivative and the Lie commutator, 
which is usually represented by a finite dimensional 
matrix commutator, share
the same Lie algebraic properties at the abstract level,
this observation suggests that the Lie derivative, 
associated with diff$N_{2}$ here, may be viewed 
as a natural infinite dimensional generalization of the finite 
dimensional matrix commutator. 
The appearance of an infinite dimensional symmetry
such as diff$N_{2}$ is not that surprising, 
since in general relativity the underlying symmetry is 
the infinite dimensional group of
diffeomorphisms of 4-dimensional spacetime.
It is the diff$N_{2}$ symmetry, the subgroup of diffeomorphisms 
of 4-dimensional spacetime, that shows up as a local 
gauge symmetry\footnote{Here local 
means local in the (1+1)-dimensional base manifold $M_{1+1}$.} of 
the Yang-Mills type, where the usual minimal couplings are 
replaced by the Lie derivatives along the vector fields $A_{\mu}$.
We also note that $\phi_{a b}$ and $\gamma_{\mu\nu}$ transform as 
a tensor field and scalar 
fields under the diff$N_{2}$ transformations, 
respectively, as can be seen from (\ref{var2}) and (\ref{var3}).
This implies that 
the (2,2)-KK formalism can be made a viable method 
of studying general relativity
from the standpoint of (1+1)-dimensional gauge theories
with an infinite dimensional gauge symmetry of the Yang-Mills type.

The prescription to obtain the scalar curvature of 
4-dimensional spacetime 
in terms of the KK variables is given in Appendix \ref{a2}.
Here we present the result only. It is given by
\begin{eqnarray}
R&=&\gamma^{\mu\nu}\tilde{R}_{\mu\nu} + \phi^{a c}R_{a c}
 + {1\over 4}\gamma^{\mu\nu}\gamma^{\alpha\beta}
  \phi_{a b}F_{\mu\alpha} ^ { \  \ a}
  F_{\nu\beta}^{\ \ b}
+{1\over 4}\gamma^{\mu\nu}\phi ^ {a b}\phi ^ {c d}\Big\{
  (D_{\mu}\phi_{a c})(D_{\nu}\phi_{b d})\nonumber\\
& & -(D_{\mu}\phi_{a b})(D_{\nu}\phi_{c d}) \Big\}   
+{1\over 4}\phi ^ {a b}\gamma^{\mu\nu}
  \gamma^{\alpha\beta}\Big\{
 (\partial_{a}\gamma_{\mu \alpha})(\partial_{b}\gamma_{\nu\beta})
 -(\partial_{a}\gamma_{\mu \nu})(\partial_{b}
  \gamma_{\alpha\beta}) \Big\}          \nonumber\\
& & +(\hat{\nabla}_{\mu}+\hat{\Gamma}_{c \mu}^{\ \ c})j^{\mu}
 +(\hat{\nabla}_{a}
 +\hat{\Gamma}_{\alpha a}^{\ \ \alpha})j^{a},  \label{too}
\end{eqnarray}
using the notations defined in Appendix \ref{a2}.
The Einstein-Hilbert action is given by the integral 
\begin{equation}
S=\int \! \! d^{2}x  d^{2}y \,
\sqrt{-\gamma}\sqrt{\phi} \, R,           \label{eh}
\end{equation}
where $\gamma={\rm det} \, \gamma_{\mu\nu}$ and
$\phi={\rm det} \, \phi_{a b}$. 
One can easily recognize that this action 
is in a form of a (1+1)-dimensional field theory action.
In geometrical terms the above action can be understood 
as follows. 
First, $\gamma^{\mu\nu}\tilde{R}_{\mu\nu}$
is the scalar curvature\footnote{
It may be worth noting that,
from the standpoint of the (1+1)-dimensional base manifold $M_{1+1}$,
the scalar curvature of the horizontal space
looks like a ``gauged'' scalar curvature in that it has
the diff$N_{2}$-valued gauge fields coupled to
$\gamma_{\mu\nu}$ in the curvature formula (\ref{gric}).} 
of the horizontal 2-space normal to the fibre space $N_{2}$.
Second, $\phi^{a c}R_{a c}$ is the scalar curvature of $N_{2}$. 
When integrated over $N_{2}$, 
it becomes a number proportional to the Euler characteristics 
of the fibre space $N_{2}$. Namely, by the Gauss-Bonnet theorem, 
we have
\begin{equation}
\int \! \! d^{2}y \, \sqrt{\phi} \, \phi^{a c}R_{a c}
=-4\pi \chi,                             \label{euler}
\end{equation}
where $\chi$ is the Euler characteristics\footnote{
Note that 
the extra minus sign in (\ref{euler}) is due to our sign convention
of curvature formula, which is the minus of more conventional 
one\cite{mtw}. For instance $\chi$ is 2 for $S^{2}$.}.
Third, apart from the last two terms containing 
$j^{\mu}$ and $j^{a}$, which are surface terms 
as is shown in Appendix \ref{a2}, the remaining terms in (\ref{too}) 
are the {\it extrinsic} terms, 
telling how the two 2-surfaces are embedded into the
enveloping 4-dimensional spacetime. They are
the Yang-Mills term, and a generic non-linear sigma model 
and a scalar field term.
Thus, the Einstein-Hilbert 
action is describable as a (1+1)-dimensional
Yang-Mills type gauge theory coupled to 
a (1+1)-dimensional non-linear 
sigma field and scalar fields of generic types, 
with intrinsic curvature terms of the two 2-surfaces. 
The associated Yang-Mills gauge symmetry is the group of 
diffeomorphisms of $N_{2}$.

Although the action (\ref{eh}) already has several 
features of gauge theories as described above, 
it appears rather complicated that its practical value 
is a bit obscure.  In the following section, we will find 
a simple action integral, 
more faithful to the spirit of gauge theories, 
by suitably gauge-fixing some of the spacetime diffeomorphisms.
This gauge-fixing brings out the Yang-Mills aspects of 
general relativity more clearly, as we shall see shortly.
\end{section}

\begin{section}{Kaluza-Klein Formalism in Polyakov Gauge}
\label{s3}

In the previous section we presented the Einstein-Hilbert action 
in the general $(2,2)$-splitting of spacetimes in terms of 
the KK variables. The most noticeable features of
the above formalism, among others, are that 
the diff$N_{2}$ symmetry appears as the {\it built-in} 
local gauge symmetry 
of the Yang-Mills type and that $A_{\mu}^{\ a}$ can be identified as 
the associated gauge fields. 
Let us recall that we can always gauge away two functions 
in the metric $\gamma_{\mu\nu}$ by a suitable
coordinate transformation, so that the metric can be 
written as, at least locally,
\begin{eqnarray}
\gamma_{\mu\nu}=\left( \begin{array}{rr}
                       -2h & -1 \\
                       -1  &  0 
\end{array}\right), \hspace{1cm} 
\gamma^{\mu\nu}=\left( \begin{array}{rr}
                        0 &  -1 \\
                       -1  &  2h 
\end{array}\right).        \label{poly}
\end{eqnarray}
Let ($u$, $v$) be a new coordinate system on $M_{1+1}$, 
in which the metric $\gamma_{\mu\nu}$ is of the form (\ref{poly}).
In the coordinate ($u$, $v$, $y^{a}$), 
the line element (\ref{gen}) in section \ref{s2} becomes 
\begin{equation}
ds^{2} =- 2 \, du dv -2 h \, du^{2}+ \phi_{a b} 
  \Big( dy^{a} + A_{+}^{\ a} du + A_{-}^{\ a} dv \Big) 
\Big( dy^{b} + A_{+}^{\ b} du + A_{-}^{\ b} dv \Big).   \label{res}
\end{equation}
It is easy to see that the remaining fields 
$\phi_{a b}$, $A_{\pm}^{\ a}$, and $h$ still transform
as a tensor field, gauge fields, and a scalar field, respectively,
under the diff$N_{2}$ transformations. Under the diff$N_{2}$ 
transformations 
\begin{equation}
y^{' a}=y^{' a} (u,v,y), \hspace{1cm} u'=u,  \hspace{1cm} 
v'=v,                                 \label{coor}
\end{equation}
we find that
$\phi_{a b}$, $A_{\pm}^{\ a}$, and $h$ transform as
\begin{eqnarray}
& &\phi'_{a b}(u',v',y')={\partial y^{c} \over \partial y^{' a}}
   {\partial y^{d} \over \partial y^{' b}}
   \phi_{c d}(u,v,y),               \label{trane} \\ 
& &A_{\pm}^{\ 'a}(u',v',y')={\partial y^{' a} \over \partial y^{c}}
   A_{\pm}^{\ c}(u,v,y) -\partial_{\pm} y^{' a}(u,v,y),  
                                            \label{trand}\\ 
& &h'(u',v',y')=h(u,v,y),                       \label{tran}
\end{eqnarray}
where $\partial_{+}=\partial / \partial u$ and 
$\partial_{-}=\partial / \partial v$. 
Under the corresponding infinitesimal transformations
\begin{eqnarray}
& &\delta y^{ a}=\xi^{a} (u,v,y) 
\hspace{1cm} ({\rm O}(\xi^{2}) \ll 1), \nonumber\\
& &\delta u=\delta v=0, 
\end{eqnarray}
they transform as 
\begin{eqnarray}
& &\delta \phi_{a b}(u,v,y)
  =-[\xi,  \phi]_{ {\rm L} a b}, \label{tranc}\\
& &\delta A_{\pm}^{\ a}(u,v,y)=-D_{\pm}\xi^{a}
   =-\partial_{\pm}\xi^{a} 
  + [A_{\pm}, \xi]_{\rm L}^{a}, \label{tranb}\\
& &\delta h(u,v,y)= -[\xi, h]_{\rm L}.             \label{trana}
\end{eqnarray}
Here the Lie derivatives are defined as before, 
\begin{eqnarray}
& &[\xi,  \phi]_{ {\rm L}a b}=\xi^{c}\partial_{c}\phi_{a b}
   +(\partial_{a}\xi^{c})\phi_{c b}
   +(\partial_{b}\xi^{c})\phi_{a c},             \\
& &[A_{\pm}, \xi]_{\rm L}^{a}=A_{\pm}^{\ c}\partial_{c}\xi^{a}
   - \xi^{c}\partial_{c}A_{\pm}^{\ a},          \\
& &[\xi,  h]_{\rm L}= \xi^{c}\partial_{c} h.
\end{eqnarray}
This observation tells us two things. First, diff$N_{2}$ is 
(a part of) the residual symmetry of the line 
element (\ref{res}), and the presence of this residual
symmetry allows us to reduce the metric further.
To see this, let us consider the following infinitesimal 
transformation of the gauge field $A_{-}^{\ a}$,
\begin{equation}
A_{-}^{\ 'a}=A_{-}^{\ a}-\partial_{-}\xi^{a} 
+ [A_{-}, \xi]_{\rm L}^{a}. 
\end{equation}
Let $\xi^{a}$ be a solution of the equation $A_{-}^{\ 'a}=0$. 
This equation defines a system of the first-order 
linear partial differential 
equations\cite{diff} for $\xi^{a}$. 
The existence of a solution of the above equation
implies that we can always choose the gauge condition
\begin{equation} 
A_{-}^{\ 'a}=0,                     \label{good}
\end{equation}
at least locally. When this condition is satisfied, 
the line element can be written as
\begin{equation}
ds^{2} =- 2 \, du dv -2 h \, du^{2}+ \phi_{a b}
  \Big( dy^{a} + A_{+}^{\ a} du \Big)
\Big( dy^{b} + A_{+}^{\ b} du \Big),   \label{resa}
\end{equation}
and its inverse metric is given by
\begin{eqnarray}
g^{AB}=\left(
\begin{array}{ccc}
0 & -1 & 0 \\
-1 & 2h & A_{+}^{\ a}\\
0 & A_{+}^{\ a} & \phi^{a b}
\end{array}\right).                  \label{fixed}
\end{eqnarray}
The coordinate system in which the metric is 
of the form (\ref{resa})
shall be referred to as the Newman-Unti coordinate\cite{unti},
for reasons that will become clear in section \ref{s5}.
Here we simply note that 
the $u={\rm constant}$ surface in (\ref{resa})
is a degenerate null hypersurface. 
The null vector field that lies in the null hypersurface 
is $\partial / \partial v$, since it has a zero norm. 
In each null hypersurface
labeled by $u$, the $v={\rm constant}$ 
section defines a 2-dimensional 
spacelike surface $N_{2}$ transverse to the vector field 
$\partial / \partial v$. 
In order to exhibit the diff$N_{2}$ symmetry manifest, however, 
we shall work mostly with the line element (\ref{res})
without picking up the condition (\ref{good}),
although we can always enforce this condition
whenever necessary. 
Second, it is clear that diff$N_{2}$ can be still viewed as the
local gauge symmetry of the Yang-Mills type, 
in view of the transformation properties of the fields.

The scalar curvature of 
4-dimensional spacetime in the Polyakov gauge (\ref{poly}) 
can be written in a remarkably simple form. 
In order to see this, let us compute each term
in (\ref{too}) in section \ref{s2} using the Polyakov metric. 
The formulae for the connection coefficients 
$\hat{\Gamma}_{\mu\nu}^{\ \ \ \alpha}$ are 
given in (\ref{com1}) in Appendix \ref{a2},
which become, in the Polyakov gauge,
\begin{eqnarray}
& &\hat{\Gamma}_{++}^{\ \ \ +}= -\hat{\Gamma}_{+-}^{\ \ \ -}
   =-\hat{\Gamma}_{-+}^{\ \ \ -} =-D_{-}h,  \nonumber\\
& &\hat{\Gamma}_{++}^{\ \ \ -}=D_{+}h + 2hD_{-}h,  \label{eval}
\end{eqnarray}
and zero otherwise. Here $D_{\pm}h$, the 
diff$N_{2}$-covariant derivatives 
of the scalar field $h$, are defined as
\begin{eqnarray}
D_{\pm}h &=& \partial_{\pm}h - [A_{\pm}, h] \nonumber\\
         &=&\partial_{\pm}h 
            -A_{\pm}^{ \ a} \partial_{a}h.    \label{eichi}
\end{eqnarray}
The Ricci tensor $\tilde{R}_{\mu\nu}$
of the horizontal 2-space is given in (\ref{gric}) 
in Appendix \ref{a2}. 
They become, in the Polyakov gauge, 
\begin{eqnarray}
& &\tilde{R}_{--}=0, \nonumber\\
& &\tilde{R}_{+-}=\tilde{R}_{-+}=-D_{-}^{2}h, \nonumber\\
& &\tilde{R}_{++}=D_{+}D_{-}h - D_{-}D_{+}h 
     - 2 h D_{-}^{2}h                \nonumber\\
& & \hspace{.9cm}=-F_{+-}^{\ \ a}\partial_{a}h 
  - 2 h D_{-}^{2} h,               \label{gra}
\end{eqnarray}
where in the last line we used the identity
\begin{equation}
D_{+}D_{-}h- D_{-}D_{+}h=-F_{+-}^{\ \ a}\partial_{a}h, \label{bebe}
\end{equation}
where 
\begin{equation}
F_{+-}^{\ \ a}=\partial_{+} A_{-} ^ { \ a}-\partial_{-}
  A_{+} ^ { \ a} - [A_{+}, A_{-}]_{\rm L}^{a}. 
\end{equation}
Therefore we have 
\begin{eqnarray}
\gamma^{\mu\nu}\tilde{R}_{\mu\nu}
&=&2\gamma^{+-}\tilde{R}_{+-} \nonumber\\
&=&2D_{-}^{2}h.  \label{gs}
\end{eqnarray}
Let us further notice that
\begin{equation}
\gamma^{\mu\nu}\partial_{a}\gamma_{\mu\nu}
={2\over \sqrt{-\gamma}}\partial_{a}\sqrt{-\gamma}=0, \label{simpq}
\end{equation}
since $\sqrt{-\gamma}=1$. 
Also we have 
\begin{equation}
\gamma^{\mu\nu}\gamma^{\alpha\beta}
(\partial_{a}\gamma_{\mu \alpha})
(\partial_{b}\gamma_{\nu\beta})=0,        \label{simpr}
\end{equation}
since $\gamma^{++}=0$ and $\gamma^{+-}=-1$.
Therefore it follows that
\begin{equation}
{1\over 4}\phi ^ {a b}\gamma^{\mu\nu}
  \gamma^{\alpha\beta}\Big\{
 (\partial_{a}\gamma_{\mu \alpha})(\partial_{b}\gamma_{\nu\beta})
 -(\partial_{a}\gamma_{\mu \nu})(\partial_{b}
  \gamma_{\alpha\beta}) \Big\}=0,    \label{quad}
\end{equation}
provided that the Polyakov gauge is chosen.
The terms containing $D_{\pm}\phi_{a b}$ must also
be written in this gauge.
The diff$N_{2}$-covariant derivative of $\phi_{a b}$ 
in the Polyakov gauge is defined as
\begin{eqnarray}
D_{\pm}\phi_{a b} &=& \partial_{\pm}\phi_{a b}
  -[A_{\pm}, \phi]_{ {\rm L} a b}      \nonumber\\
&=&\partial_{\pm}\phi_{a b}
  -A_{\pm} ^ { \ c}\partial_c \phi_{a b}
  -(\partial_a A_{\pm} ^ {\ c})\phi_{c b}
  -(\partial_b A_{\pm} ^ { \ c})\phi_{a c}. \label{cov}
\end{eqnarray}
It is well-known that, in the null hypersurface formalism of 
general relativity, the transverse 2-metric with a unit 
determinant, i.e. the conformal 2-metric\cite{sach,inv,inva,small}
represents the two physical 
degrees of freedom of gravitational 
field\footnote{The gauge in which the metric
is of the form (\ref{res}) may be viewed as an
analog of the Coulomb gauge in Maxwell's theory, where
the physical degrees of freedom are the transverse traceless
vector potentials, since a finite generalization of traceless 
condition is a unit determinant condition.}. 
Let us therefore 
decompose the metric $\phi_{a b}$ of 
the fibre space $N_{2}$ into a volume field 
$\phi$  and
a conformal 2-metric $\rho_{ab}$,
\begin{equation}
\phi_{a b}=\sqrt{\phi}\ \rho_{a b}= {\rm e}^{\sigma} \rho_{ab},
\end{equation}
where $\rho_{a b}$ satisfies the condition
\begin{equation}
{\rm det}\, \rho_{a b} = 1,        \label{confor}
\end{equation}
so that $\phi={\rm det}\, \phi_{a b}$. 
The field $\sigma$ is related to $\phi$ by the equation
\begin{equation}
\sigma ={1\over 2}\, {\rm ln}\, \phi.       \label{newsig}
\end{equation}
Notice that $\rho_{a b}$ is a tensor density 
with the weight $+1$, and that $\phi$ is a scalar density 
with the weight $-2$ with respect to the diff$N_{2}$ transformation, 
respectively. Therefore the diff$N_{2}$-covariant 
derivatives\footnote{In general, 
for the tensor density $T_{ab\cdots}$ with the weight $w$ 
with respect to the diff$N_{2}$ transformations, 
we have $D_{\pm}T_{ab\cdots}
=\partial_{\pm}T_{ab\cdots}-[A_{\pm}, T]_{{\rm L}ab\cdots}$,
which includes the term 
$+w(\partial_{c}A_{\pm}^{\ c})T_{ab\cdots}$.}
of $\rho_{ab}$ and $\phi$ are naturally given by
\begin{eqnarray}
& &D_{\pm}\rho_{a b}=\partial_{\pm}\rho_{a b}
   - [A_{\pm}, \rho]_{{\rm L}a b} \nonumber\\
& &\hspace{1.25cm}=\partial_{\pm}\rho_{a b}
-A_{\pm} ^ { \ c}\partial_c \rho_{a b}
-(\partial_a A_{\pm} ^ { \ c})\rho_{c b}
-(\partial_b A_{\pm} ^ { \ c})\rho_{a c}
+(\partial_c A_{\pm} ^ { \ c})\rho_{a b},  \label{rhod}\\
& &D_{\pm}\phi =\partial_{\pm}\phi 
  -[A_{\pm}, \phi]_{\rm L}           \nonumber\\
& &\hspace{.95cm}=\partial_{\pm}\phi
    -A_{\pm}^{\ a}\partial_{a}\phi
    -2(\partial_{a}A_{\pm}^{\ a})\phi,     \label{densig}
\end{eqnarray}
respectively. The diff$N_{2}$-covariant 
of $\sigma$ follows automatically from (\ref{densig}), 
\begin{eqnarray}
& &D_{\pm}\sigma = \partial_{\pm}\sigma
-[A_{\pm}, \sigma]_{\rm L}        \nonumber\\
& &\hspace{.95cm}=\partial_{\pm}\sigma
-A_{\pm}^{\ a}\partial_{a}\sigma
-\partial_{a}A_{\pm}^{\ a}.         \label{get}
\end{eqnarray}
Using the identity 
\begin{equation}
\rho^{a b}D_{\pm}\rho_{a b}=0,                 \label{ident}
\end{equation}
which is as a trivial consequence 
of (\ref{confor}) and (\ref{rhod}), we find that
\begin{eqnarray}
\phi^{a b}D_{\pm}\phi_{a b}
&=&{\rm e}^{-\sigma}\rho^{a b}
\Big\{ {\rm e}^{\sigma}(D_{\pm}\sigma )\rho_{a b}
+ {\rm e}^{\sigma}D_{\pm}\rho_{a b} \Big\} \nonumber\\
&=&2D_{\pm}\sigma.                            \label{extr}
\end{eqnarray}
Let us notice that we also have
\begin{eqnarray}
& &\phi^{a b}\phi^{c d}(D_{\pm}\phi_{a c})
  (D_{-}\phi_{b d})               \nonumber\\
&=&{\rm e}^{-2\sigma}\rho^{a b}\rho^{c d}\Big\{
{\rm e}^{\sigma} (D_{\pm}\sigma )\rho_{a c}
+{\rm e}^{\sigma}(D_{\pm}\rho_{a c}) \Big\}
\Big\{
{\rm e}^{\sigma}  (D_{-}\sigma )\rho_{b d}
+{\rm e}^{\sigma} (D_{-}\rho_{b d} )\Big\}  \nonumber\\
&=&2(D_{\pm}\sigma )(D_{-}\sigma )
 +\rho^{a b} (D_{\pm}\sigma )(D_{-}\rho_{a b}) 
 +\rho^{a b}(D_{-}\sigma ) (D_{\pm}\rho_{a b})
 +\rho^{a b}\rho^{c d}
 (D_{\pm}\rho_{a c})(D_{-}\rho_{b d})     \nonumber\\
&=&2(D_{\pm}\sigma )(D_{-}\sigma )
 +\rho^{a b}\rho^{c d}
 (D_{\pm}\rho_{a c})(D_{-}\rho_{b d}),  \label{enla}
\end{eqnarray}
using the equation (\ref{ident}).
Therefore the term containing $D_{\pm}\phi_{a b}$ becomes
\begin{eqnarray}
& &{1\over 4}\gamma^{\mu\nu}\phi ^ {a b}\phi ^ {c d}\Big\{
  (D_{\mu}\phi_{a c})(D_{\nu}\phi_{b d})
 -(D_{\mu}\phi_{a b})(D_{\nu}\phi_{c d}) \Big\}  \nonumber\\
&=&(D_{+}\sigma)(D_{-}\sigma)   
  -{1\over 2}\rho^{a b}\rho^{c d}
  (D_{+}\rho_{a c})(D_{-}\rho_{b d})  
  -h\Big\{  (D_{-}\sigma)^{2}       
-{1\over 2}\rho^{a b}\rho^{c d}
 (D_{-}\rho_{a c})(D_{-}\rho_{b d})\Big\}.  \label{enlb}
\end{eqnarray}
Also the Yang-Mills type term becomes,
\begin{equation}
{1\over 4}\gamma^{\mu\nu}\gamma^{\alpha\beta}
  \phi_{a b}F_{\mu\alpha} ^ { \  \ a}
  F_{\nu\beta}^{\ \ b}
= - {1\over 2}{\rm e}^{\sigma}\rho_{a b}
  F_{+-}^{\ \ a}F_{+-}^{\ \ b}.           \label{ym}
\end{equation}
The remaining terms are the
surface terms when integrated over the spacetime 
coordinates, as is shown in Appendix \ref{a2}.
For completeness, we shall evaluate them 
in the Polyakov gauge, using the defining equations of 
$j^{\pm}$ and $j^{a}$, 
\begin{eqnarray}
& &j^{\mu}=\gamma^{\mu\nu}
        \phi^{a b}D_{\nu}\phi_{a b},           \label{ja}\\
& &j^{a}=\phi^{a b} \gamma^{\mu\nu}
        \partial_{b}\gamma_{\mu\nu},           \label{je}
\end{eqnarray}
as is given in (\ref{jeimu}) and (\ref{jeiei}) 
in Appendix \ref{a2}. From the equation (\ref{simpq}) 
we have
\begin{equation}
j^{a}=0,
\end{equation}
and $j^{\pm}$ are found to be
\begin{eqnarray}
& &j^{+}=-2 D_{-}\sigma,             \label{jeia}\\
& &j^{-}=-2 D_{+}\sigma + 4h D_{-}\sigma. \label{jeib}
\end{eqnarray}
Therefore the surface term containing $j^{\pm}$
becomes, 
\begin{eqnarray}
\Big( \hat{\nabla}_{\mu}
+\hat{\Gamma}_{c \mu}^{\ \ c} \Big) j^{\mu}
&=&\hat{\partial}_{\mu}j^{\mu}
   +\hat{\Gamma}_{\mu\nu}^{\ \ \mu}j^{\nu}
   +\hat{\Gamma}_{c \mu}^{\ \ c}j^{\mu}       \nonumber\\
&=&\hat{\partial}_{+} j^{+} +\hat{\partial}_{-} j^{-}
+\hat{\Gamma}_{+ +}^{\ \ \ +} j^{+} 
+\hat{\Gamma}_{- +}^{\ \ \ -} j^{+}
+\hat{\Gamma}_{c +}^{\ \ c} j^{+}
+\hat{\Gamma}_{c -}^{\ \ c} j^{-}  \nonumber\\
&=&\Big( 
\hat{\partial}_{+} + 
\hat{\Gamma}_{c +}^{\ \ c} \Big)j^{+}
+\Big( \hat{\partial}_{-} + \hat{\Gamma}_{c -}^{\ \ c}
   \Big)j^{-},              \label{curr}
\end{eqnarray}
where we used the identities 
\begin{eqnarray}
& &\hat{\Gamma}_{+ -}^{\ \ \ +}=\hat{\Gamma}_{- -}^{\ \ \ -}=0, 
   \nonumber\\
& &\hat{\Gamma}_{+ +}^{\ \ \ +}
=-\hat{\Gamma}_{- +}^{\ \ \ -}=-D_{-}h.
\end{eqnarray}
Using the equation
\begin{equation}
\hat{\Gamma}_{c \pm}^{\ \ c}=D_{\pm}\sigma,
\end{equation}
which follows from (\ref{contrad}) in Appendix \ref{a2} and
(\ref{extr}), the surface term finally becomes,
\begin{eqnarray}
\Big( \hat{\nabla}_{\mu}+\hat{\Gamma}_{c \mu}^{\ \ c}\Big) j^{\mu}
&=&-2\Big( \partial_{+} - A_{+}^{\ a}\partial_{a}
+D_{+}\sigma \Big) ( D_{-}\sigma )  \nonumber\\
& &-2\Big(
\partial_{-} - A_{-}^{\ a}\partial_{a}
+D_{-}\sigma \Big)
\Big(  D_{+}\sigma - 2 h D_{-}\sigma \Big).        \label{ham}
\end{eqnarray}
Thus, putting the equations (\ref{gs}), (\ref{quad}),
(\ref{enlb}), (\ref{ym}), and (\ref{ham}) all together into
(\ref{too}) in section \ref{s2}, 
we obtain the scalar curvature\footnote{
Here the subscript ${}_{\rm p}$ means that the quantity
was evaluated in the Polyakov gauge.} $R_{\rm p}$
of 4-dimensional spacetime 
in the Polyakov gauge. It is given by 
\begin{eqnarray}
R_{\rm p}&=&R_{2} + 2D_{-}^{2} h 
  -{1\over 2}{\rm e}^{\sigma}\rho_{a b}
  F_{+-}^{\ \ a}F_{+-}^{\ \ b}
  +(D_{+}\sigma) (D_{-}\sigma)       
  -{1\over 2}\rho^{a b}\rho^{c d} 
 (D_{+}\rho_{a c})(D_{-}\rho_{b d})   \nonumber\\
& & -h\Big\{ 
    (D_{-}\sigma)^{2}
    -{1\over 2}\rho^{a b}\rho^{c d} 
    (D_{-}\rho_{a c})(D_{-}\rho_{b d}) \Big\}
-2\Big(
\partial_{+} - A_{+}^{\ a}\partial_{a}
+D_{+}\sigma \Big)
( D_{-}\sigma )               \nonumber\\
& &-2\Big(
\partial_{-} - A_{-}^{\ a}\partial_{a} 
+D_{-}\sigma \Big)
\Big(D_{+}\sigma - 2h D_{-}\sigma \Big),       \label{lag} 
\end{eqnarray}
where $R_2=\phi^{a c}R_{a c}$ is the scalar curvature 
of the fibre space $N_{2}$. 
The integral of $R_{\rm p}$ over the 4-dimensional spacetime, 
which we denote by $S_{\rm p}$, is given by
\begin{equation}
S_{\rm p}=\int \! \! du \, dv \, d^{2}y \,
{\rm e}^{\sigma} R_{\rm p},  \label{laga}
\end{equation}
since $\sqrt{-\gamma}=1$.
Ignoring the surface integrals 
coming from the last two terms in $R_{\rm p}$, 
the integral $S_{\rm p}$ can be written as
\begin{eqnarray}
S_{\rm p} & = & \int \! \! du \, dv \, d^{2}y \,
\Big[ -{1\over 2}{\rm e}^{2 \sigma}\rho_{a b}
  F_{+-}^{\ \ a}F_{+-}^{\ \ b}
  +{\rm e}^{\sigma} (D_{+}\sigma) (D_{-}\sigma)  
-{1\over 2}{\rm e}^{\sigma}\rho^{a b}\rho^{c d}
 (D_{+}\rho_{a c})(D_{-}\rho_{b d})  \nonumber\\
& &+{\rm e}^{\sigma}R_2  
   + 2{\rm e}^{\sigma}D_{-}^{2} h           
   -h{\rm e}^{\sigma}\Big\{
 (D_{-}\sigma)^{2}
-{1\over 2}\rho^{a b}\rho^{c d}
 (D_{-}\rho_{a c})(D_{-}\rho_{b d})  \Big\} \Big] \nonumber\\
& &+ \ {\rm surface} \ {\rm terms}.    \label{reaction}
\end{eqnarray}
Let us now notice that the term ${\rm e}^{\sigma}D_{-}^{2} h$ 
can be written as 
\begin{eqnarray}
 {\rm e}^{\sigma}D_{-}^{2} h 
&=&-{\rm e}^{\sigma} ( D_{-}\sigma ) ( D_{-}h )
   +D_{-}\Big( {\rm e}^{\sigma} D_{-}h \Big)       \nonumber\\
&=&-{\rm e}^{\sigma} (D_{-}\sigma) (D_{-}h)
   +\partial_{-}\Big( {\rm e}^{\sigma}  D_{-}h \Big)
   -\partial_{a}\Big(  
   A_{-}^{\ a} {\rm e}^{\sigma} D_{-}h\Big). \label{del}
\end{eqnarray}
But the first term in (\ref{del}) becomes
\begin{eqnarray}
& &-{\rm e}^{\sigma} (D_{-}\sigma) (D_{-}h)   \nonumber\\
&=&hD_{-}\big(   {\rm e}^{\sigma}  D_{-}\sigma    \Big)
  -D_{-} \Big( h {\rm e}^{\sigma}D_{-}\sigma \Big) \nonumber\\
&=&h{\rm e}^{\sigma}(D_{-}^{2}\sigma)
+h{\rm e}^{\sigma}(D_{-}\sigma)^{2} 
-\partial_{-}\Big( h {\rm e}^{\sigma}D_{-}\sigma \Big)
+\partial_{a}\Big( 
hA_{-}^{\ a}{\rm e}^{\sigma} D_{-}\sigma \Big). \label{dela}
\end{eqnarray}
Therefore we find that
${\rm e}^{\sigma}D_{-}^{2} h$ becomes
\begin{eqnarray}
{\rm e}^{\sigma}D_{-}^{2} h &=&
h{\rm e}^{\sigma} ( D_{-}^{2}\sigma )
   +h{\rm e}^{\sigma}( D_{-}\sigma )^{2}
   +\partial_{-}\Big({\rm e}^{\sigma}D_{-}h  \Big)
   -\partial_{-}\Big(  h {\rm e}^{\sigma}
   D_{-}\sigma   \Big)         \nonumber\\ 
& &   -\partial_{a}\Big(
  A_{-}^{\ a} {\rm e}^{\sigma}D_{-}h \Big) 
  +\partial_{a}\Big( h  A_{-}^{\ a}{\rm e}^{\sigma}
    D_{-}\sigma \Big).                  \label{ADDB}
\end{eqnarray}
Thus the integral $S_{\rm p}$ becomes
\begin{eqnarray}
S_{\rm p}& = & \int \! \! du \, dv \, d^{2}y \,
\Big[ -{1\over 2}{\rm e}^{2 \sigma}\rho_{a b}
  F_{+-}^{\ \ a}F_{+-}^{\ \ b}
  +{\rm e}^{\sigma} (D_{+}\sigma) (D_{-}\sigma) 
  -{1\over 2}{\rm e}^{\sigma}\rho^{a b}\rho^{c d}
 (D_{+}\rho_{a c})(D_{-}\rho_{b d})  \nonumber\\
& & +{\rm e}^{\sigma} R_2       
    +2h{\rm e}^{\sigma}\Big\{
  D_{-}^{2}\sigma +{1\over 2} (D_{-}\sigma)^{2} 
 + {1\over 4}\rho^{a b}\rho^{c d}
   (D_{-}\rho_{a c})(D_{-}\rho_{b d}) \Big\}  
    \Big]                       \nonumber\\
& & \ + {\rm surface} \ {\rm terms}   \nonumber\\
&\equiv & \int \! \! du \, dv \, d^{2}y \, 
    {\cal L}_{\rm p}\ 
+ {\rm surface} \ {\rm terms},      \label{reactiona}     
\end{eqnarray}
where the integrand ${\cal L}_{\rm p}$ is defined as 
\begin{eqnarray}
{\cal L}_{\rm p}&=& -{1\over 2}{\rm e}^{2 \sigma}\rho_{a b}
  F_{+-}^{\ \ a}F_{+-}^{\ \ b}
  +{\rm e}^{\sigma} (D_{+}\sigma) (D_{-}\sigma) 
 -{1\over 2}{\rm e}^{\sigma}\rho^{a b}\rho^{c d}
 (D_{+}\rho_{a c})(D_{-}\rho_{b d})      \nonumber\\
& &+{\rm e}^{\sigma} R_{2} +2h{\rm e}^{\sigma}\Big\{
  D_{-}^{2}\sigma +{1\over 2} (D_{-}\sigma)^{2}
 + {1\over 4}\rho^{a b}\rho^{c d}
   (D_{-}\rho_{a c})(D_{-}\rho_{b d}) \Big\}. \label{elzero}
\end{eqnarray}
This Lagrangian density ${\cal L}_{\rm p}$ 
can be naturally interpreted as describing
the Yang-Mills type gauge theory defined on the (1+1)-dimensional
base manifold $M_{1+1}$, interacting with the (1+1)-dimensional 
``matter'' fields $\rho_{a b}$ and $\sigma$. 
The associated local gauge symmetry is 
the {\it built-in} diff$N_{2}$ symmetry,
the group of diffeomorphisms of the fibre space $N_{2}$.
It must be mentioned here that each term in (\ref{elzero}) is 
manifestly diff$N_{2}$-invariant, and that the $y^{a}$-dependence 
of each term is completely {\it hidden} in the Lie derivative.
In this sense we may view the fibre space $N_{2}$ as 
a kind of {\it internal} space\footnote{
The integration over the $y^{a}$-coordinates in (\ref{reactiona}) 
might be viewed as an infinite dimensional
generalization of taking finite dimensional {\it trace}
over the fibre space indices
in Yang-Mills theories with some finite dimensional gauge
symmetries.} as in Yang-Mills theory.

In order to derive the Einstein's equations as 
the Euler-Lagrange equations of motion in this KK formalism, 
the Lagrangian density ${\cal L}_{\rm p}$ must be implemented
with two equations associated with gauge-fixing of the metric 
$\gamma_{\mu\nu}$ of the horizontal 2-space to 
the Polyakov metric. Notice that the eight field equations 
are obtainable from the Lagrangian density ${\cal L}_{\rm p}$. 
In order to find the remaining two equations, 
we have to vary the general Einstein-Hilbert action 
(\ref{eh}) in section \ref{s2} with respect to 
the general functions $\gamma_{+-}$ and $\gamma_{--}$, 
and then use the Polyakov gauge condition 
$\gamma_{+-}=-1$ and $\gamma_{--}=0$ at the end of
the variations.
In the next section we shall present
the ten Einstein's equations 
without giving the details of derivations. 

\end{section}

\begin{section}{The Einstein's Equations}
\label{s4}

In this section we present the ten Einstein's equations. 
The eight equations can be obtained by varying ${\cal L}_{\rm p}$
with respect to $h$, $A_{\pm}^{\ a}$, $\sigma$, and $\rho_{a b}$
subject to a unit determinant condition. 
In order to enforce this condition,
we must add to ${\cal L}_{\rm p}$ a Lagrange multiplier term 
\begin{equation}
\lambda \Big( {\rm det}\, \rho_{a b} - 1 \Big),  \label{mul}
\end{equation}
where $\lambda$ is a Lagrange multiplier
to be determined.
The eight Einstein's equations are as follows:
\begin{eqnarray}
&(a)&\hspace{.5cm}2{\rm e}^{\sigma}(D_{-}^{2}\sigma) +
{\rm e}^{\sigma} (D_{-}\sigma)^{2}  
 + {1\over 2}{\rm e}^{\sigma}\rho^{a b}\rho^{c d} (D_{-}\rho_{a c})
    (D_{-}\rho_{b d})=0,                 \label{cpm}\\
&(b)&\hspace{.5cm}D_{-}\Big( {\rm e}^{2\sigma}
\rho_{a b}F_{+-}^{\ \ b}\Big)
- {\rm e}^{\sigma}\partial_{a}(D_{-}\sigma) 
- {1\over 2}{\rm e}^{\sigma}\rho^{b c}\rho^{d e}
    (D_{-}\rho_{b d})(\partial_{a}\rho_{c e}) 
+ \partial_{b} \Big(
{\rm e}^{ \sigma}\rho^{b c}D_{-}\rho_{a c} \Big)\nonumber\\
& &\hspace{.5cm}=0,\label{eqq}\\
&(c)&\hspace{.5cm}-D_{+}\Big( {\rm e}^{2\sigma}
\rho_{a b}F_{+-}^{\ \ b}\Big) 
-{\rm e}^{\sigma}\partial_{a} (D_{+}\sigma )
   -{1\over 2}{\rm e}^{\sigma}\rho^{b c}\rho^{d e}
   (D_{+}\rho_{b d})(\partial_{a}\rho_{c e})  \nonumber\\
& &\hspace{.5cm}+\partial_{b}\Big(  
{\rm e}^{ \sigma}\rho^{b c}D_{+}\rho_{a c} \Big)
+2h{\rm e}^{\sigma}\partial_{a}(D_{-}\sigma)   
+h{\rm e}^{\sigma}\rho^{b c}\rho^{d e}
   (D_{-}\rho_{b d})(\partial_{a}\rho_{c e}) 
+2{\rm e}^{\sigma}\partial_{a}(D_{-}h)    \nonumber\\
& &\hspace{.5cm}-2\partial_{b}\Big(h  
{\rm e}^{ \sigma}\rho^{b c}D_{-}\rho_{a c} \Big)
=0,                              \label{eqa}\\
&(d)&\hspace{.5cm}-2 {\rm e}^{ \sigma}D_{-}^{2} h 
-2{\rm e}^{ \sigma} (D_{-}h)(D_{-}\sigma) 
+ {\rm e}^{ \sigma}D_{+}D_{-}\sigma 
+ {\rm e}^{ \sigma}D_{-}D_{+}\sigma        
+ {\rm e}^{ \sigma} (D_{+}\sigma)(D_{-}\sigma) \nonumber\\  
& &\hspace{.5cm}+ {1\over 2}{\rm e}^{ \sigma}\rho^{a b}\rho^{c d} 
  (D_{+}\rho_{a c})(D_{-}\rho_{b d})   
+ {\rm e}^{2 \sigma}\rho_{a b}
    F_{+-}^{\ \ a}F_{+-}^{\ \ b}
 -2h{\rm e}^{ \sigma} \Big\{
   D_{-}^{2} \sigma +{1\over 2}(D_{-}\sigma)^{2} \nonumber\\
& &\hspace{.5cm}+{1\over 4}\rho^{a b}\rho^{c d}
   (D_{-}\rho_{a c})(D_{-}\rho_{b d})\Big\}=0,  \label{eqb}\\
&(e)&\hspace{.5cm}h\Big\{ {\rm e}^{\sigma} D_{-}^{2} \rho_{ab} 
- {\rm e}^{\sigma}\rho^{c d}(D_{-}\rho_{a c})(D_{-}\rho_{b d}) 
+{\rm e}^{\sigma}(D_{-}\rho_{a b})(D_{-}\sigma) \Big\}  \nonumber\\
& &\hspace{.5cm}-{1\over 2}{\rm e}^{\sigma} \Big( 
D_{+}D_{-}\rho_{a b} + D_{-}D_{+}\rho_{a b} \Big) 
+{1\over 2}{\rm e}^{\sigma} \rho^{c d}\Big\{ 
(D_{-}\rho_{a c})(D_{+}\rho_{b d}) 
+(D_{-}\rho_{b c})(D_{+}\rho_{a d}) \Big\}  \nonumber\\
& &\hspace{.5cm}-{1\over 2}{\rm e}^{\sigma}\Big\{ 
(D_{-}\rho_{a b})(D_{+}\sigma)
+(D_{+}\rho_{a b})(D_{-}\sigma)  \Big\}  \nonumber\\
& &\hspace{.5cm}+{\rm e}^{\sigma}(D_{-}\rho_{a b})(D_{-}h)                
+{1\over 2}{\rm e}^{2 \sigma}\rho_{a c}\rho_{b d}
 F_{+-}^{\ \ c}F_{+-}^{\ \ d}     
-{1\over 4}{\rm e}^{2 \sigma}\rho_{a b}
 \rho_{c d}F_{+-}^{\ \ c}F_{+-}^{\ \ d}=0. \label{prop}
\end{eqnarray}
The remaining two equations associated with the variations 
with respect to $\gamma_{+-}$ and $\gamma_{--}$ are found 
as follows:
\begin{eqnarray}
&(f)&\hspace{.5cm} {\rm e}^{\sigma} D_{+}D_{-}\sigma 
+ {\rm e}^{\sigma} D_{-}D_{+}\sigma  
+ 2{\rm e}^{\sigma} (D_{+}\sigma)(D_{-}\sigma)
- 2{\rm e}^{\sigma}(D_{-}h)(D_{-}\sigma)  \nonumber\\
& &\hspace{.5cm} - {1\over 2}{\rm e}^{ 2 \sigma}\rho_{a b}
   F_{+-}^{\ \ a}F_{+-}^{\ \ b}
- {\rm e}^{\sigma} R_{2}
- h {\rm e}^{\sigma} \Big\{
(D_{-}\sigma)^{2} 
-{1\over 2}\rho^{a b}\rho^{c d} 
 (D_{-}\rho_{a c})(D_{-}\rho_{b d})\Big\}=0, \label{eqd}\\
&(g)&\hspace{.5cm}-{\rm e}^{\sigma} D_{+}^{2}\sigma 
- {1\over 2}{\rm e}^{\sigma}(D_{+}\sigma)^{2}
-{\rm e}^{\sigma}(D_{-}h) (D_{+}\sigma)
+{\rm e}^{\sigma}(D_{+}h)(D_{-}\sigma) \nonumber\\
& &\hspace{.5cm}+2h {\rm e}^{\sigma}(D_{-}h)(D_{-}\sigma) 
+{\rm e}^{\sigma}F_{+-}^{\ \ a}\partial_{a}h
-{1\over 4}{\rm e}^{\sigma}\rho^{a b}\rho^{c d} 
 (D_{+}\rho_{a c})(D_{+}\rho_{b d})
+\partial_{a}\Big( \rho^{a b}\partial_{b}h \Big) \nonumber\\
& &\hspace{.5cm}+h\Big\{ - {\rm e}^{\sigma} 
(D_{+}\sigma) (D_{-}\sigma)
+{1\over 2}{\rm e}^{\sigma}\rho^{a b}\rho^{c d} (D_{+}\rho_{a c})
    (D_{-}\rho_{b d})
+{1\over 2}{\rm e}^{2\sigma}\rho_{a b}F_{+-}^{\ \ a}F_{+-}^{\ \ b}
+{\rm e}^{\sigma}R_{2} \Big\} \nonumber\\
& &\hspace{.5cm}+h^{2}{\rm e}^{\sigma}\Big\{
(D_{-}\sigma)^{2}
-{1\over 2}\rho^{a b}\rho^{c d}
(D_{-}\rho_{a c}) (D_{-}\rho_{b d})\Big\}=0.   \label{eqc}
\end{eqnarray}
The equations $(\ref{cpm})$, $(\ref{eqq})$, $(\ref{eqa})$, 
$(\ref{eqb})$, and $(\ref{prop})$ are the Euler-Lagrange 
equations of motion corresponding to the variations 
in $h$, $A_{+}^{\ a}$, $A_{-}^{\ a}$, $\sigma$, and
$\rho_{a b}$, respectively, and the equations 
$(\ref{eqd})$ and $(\ref{eqc})$
are the equations of motion of $\gamma_{+-}$ and $\gamma_{--}$, 
respectively. 
An inspection of these equations reveals that 
the four equations $(\ref{eqa})$, $(\ref{eqd})$, and $(\ref{eqc})$ 
do {\it not} contain second-order derivative terms in
$v$. 
This suggests that the affine parameter $v$ may be identified as the 
{\it natural} time coordinate in this formalism, and that
these four equations are the constraint equations restricting the 
initial value data on the $v={\rm constant}$ hypersurface. 
Since these four equations are written in the Polyakov gauge, they
are expected to generate the residual symmetry\footnote{
For the residual symmetry of this kind,
see the discussions in \cite{newman+tod} for instance.}\cite{solo}
of the line element (\ref{res}), i.e. the symmetry 
that survives after the Polyakov gauge is chosen.
Thus, the equations $(\ref{eqa})$ generate 
the diff$N_{2}$ symmetry, and the equations $(\ref{eqd})$ and 
$(\ref{eqc})$ generate the residual symmetry associated with 
the reparametrization invariance of 
the (1+1)-dimensional base manifold $M_{1+1}$.

The remaining equations $(\ref{cpm})$, $(\ref{eqq})$, 
$(\ref{eqb})$ and $(\ref{prop})$ are the evolution equations
since they contain the second-order $v$-time derivatives.
Notice that the equation $(\ref{cpm})$ can be 
written in a form of the Schr\"{o}dinger equation.
Using the identity
\begin{equation}
2\ D_{-}^{2}{\rm e}^{ \sigma / 2}
={\rm e}^{ \sigma /  2}\Big\{ D_{-}^{2}\sigma 
 + {1\over 2}( D_{-}\sigma)^{2}\Big\},     \label{sig}
\end{equation}
the equation (\ref{cpm}) becomes
\begin{equation}
D_{-}^{2}\psi + \kappa^{2} \psi =0,               \label{geq}
\end{equation}
where 
\begin{eqnarray}
& &\psi ={\rm e}^{ \sigma / 2},      \label{added}\nonumber\\
& &\kappa^{2}={1\over 8}\rho^{a b}\rho^{c d} 
(D_{-}\rho_{a c})(D_{-}\rho_{b d}) \geq 0. \label{kappa}
\end{eqnarray}
In Appendix \ref{a3} 
it is shown that $\kappa^{2}$ is positive definite.
It is a bilinear combination of the {\it gravitational} 
{\it current} $j^{a}_{ \ b}$,
\begin{equation}
j^{a}_{ \ b}=\rho^{a c}D_{-}\rho_{c b},       \label{cur}
\end{equation}
formed by the conformal 2-metric $\rho_{a b}$ and 
its first $v$-time derivative.
Since the true gravitational degrees of freedom are known to
reside in the conformal 2-metric $\rho_{a b}$, 
we expect that it is the 
gravitational current $j^{a}_{ \ b}$ that transports 
the true gravitational degrees of freedom. 
More often than not, important physical quantities appear 
in bilinear combinations of more fundamental quantities. 
Thus, it is of interest to find what physical quantity $\kappa^{2}$ 
represents, being a bilinear in the ``fundamental'' gravitational 
current\footnote{In the case of the Brill wave equation, 
this quantity represents a kind of gravitational energies
associated with gravitational waves.}. 

The equation (\ref{geq}) is an analog of 
the Brill wave equation\footnote{
Although the equation 
(\ref{geq}) is a three dimensional P.D.E.,
it can be made a one dimensional O.D.E. if the 
gauge condition $A_{-}^{\ a}=0$ is chosen.
Unlike the Brill wave equation, 
the wave function ${\rm e}^{ \sigma / 2}$ here is 
the square root of the conformal factor of the two dimensional
wavefront $N_{2}$, the spatial projection of the null 
hypersurface $u={\rm constant}$.}\cite{brill}, 
as it is of the type of the Schr\"{o}dinger equation
for a wave function corresponding to a state of 
zero energy in the potential $-\kappa^{2}$.
Thus $\sigma$ is a function that can be determined 
by the potential $-\kappa^{2}$ up to
some integral ``constant" function.
The generic behavior of solutions 
of the equation is therefore expected either of
the scattering type, or of the bound-state or resonance type,
corresponding to the asymptotically flat spacetimes or
spatially closed universes, respectively,
on a par with the Brill wave equation. 
\end{section}

\begin{section}{The Null Hypersurface Formalism}
\label{s5}

In this section we shall find that the KK formalism presented
in previous sections  
is in fact equivalent to the null hypersurface formalism.
This becomes quite obvious if we recognize 
that the coordinate system in which the Polyakov gauge 
condition and the extra gauge condition $A_{-}^{\ a}=0$ are 
simultaneously valid is precisely the Newman-Unti 
coordinate\cite{unti}, which is usually adopted 
in the null hypersurface formalism. We shall begin our discussions
by reviewing how the Newman-Unti coordinate is constructed
in the null hypersurface formalism.
Let us introduce a null vector field $l_{A}$, 
which we may 
choose to be the minus of the gradient of some scalar function $u$, 
\begin{equation}
l_{A}=-\nabla_{A}u.
\end{equation}
From this we can define 
another scalar function $v$ by the equation
\begin{equation}
l^{A}{\partial v \over \partial x^{A}} = 1.
\end{equation}
Solving this equation for $l^{A}$, we find that
\begin{equation}
l^{A}={\partial x^{A}\over \partial v}, 
\end{equation}
which tells us that $v$ is the (affine) parameter 
of the null curve generated by $l^{A}$.
Let us choose $x^{+}=u$,  $x^{-}=v$, and $x^{a}=y^{a} (a=2,3)$
as our coordinates. 
In this coordinate system, the components of 
$l_{A}$ and $l^{A}$ are given by
\begin{equation}
l_{A}=-\delta_{A}^{ \ +}, \hspace{1cm}
l^{A}=\delta_{-}^{\ A},            \label{els}
\end{equation}
respectively. 
From the following identity 
\begin{equation}
l^{A}=g^{A B}l_{B}=\delta_{-}^{\ A},
\end{equation}
we find that $g^{A +}$ is given by
\begin{equation}
g^{A+}=-\delta_{-}^{\ A}=(0,\ -1,\ 0, \ 0). \label{nunti}
\end{equation}
The coordinate system $(u,v, y^{a})$ in which the metric is 
of the form
(\ref{nunti}) is called the Newman-Unti coordinate, and 
the metric (\ref{fixed}) in section \ref{s3} is precisely 
of this form. 

The complete set of the null tetrads 
\begin{math}
(l,\ n, \ m, \ \bar{m})
\end{math}
is defined by the following relations
\begin{eqnarray}
& &l^{2}=n^{2}=m^{2}=\bar{m}^{2}=l \cdot m
= n \cdot m = 0,                                 \nonumber\\
& &l \cdot n = -1, \ \ \ \ m \cdot \bar{m} =1,   \label{tet}
\end{eqnarray}
where $\bar{m}$ is the complex conjugate of $m$.
In terms of these null tetrads, the metric is given by
\begin{equation}
g^{A B}=-2n^{{(}A}l^{B{)}} 
+ 2m^{{(}A}\bar{m}^{B{)}}.     \label{conditiona}
\end{equation}

\begin{figure}
\centerline{\epsfxsize=14cm\epsfbox{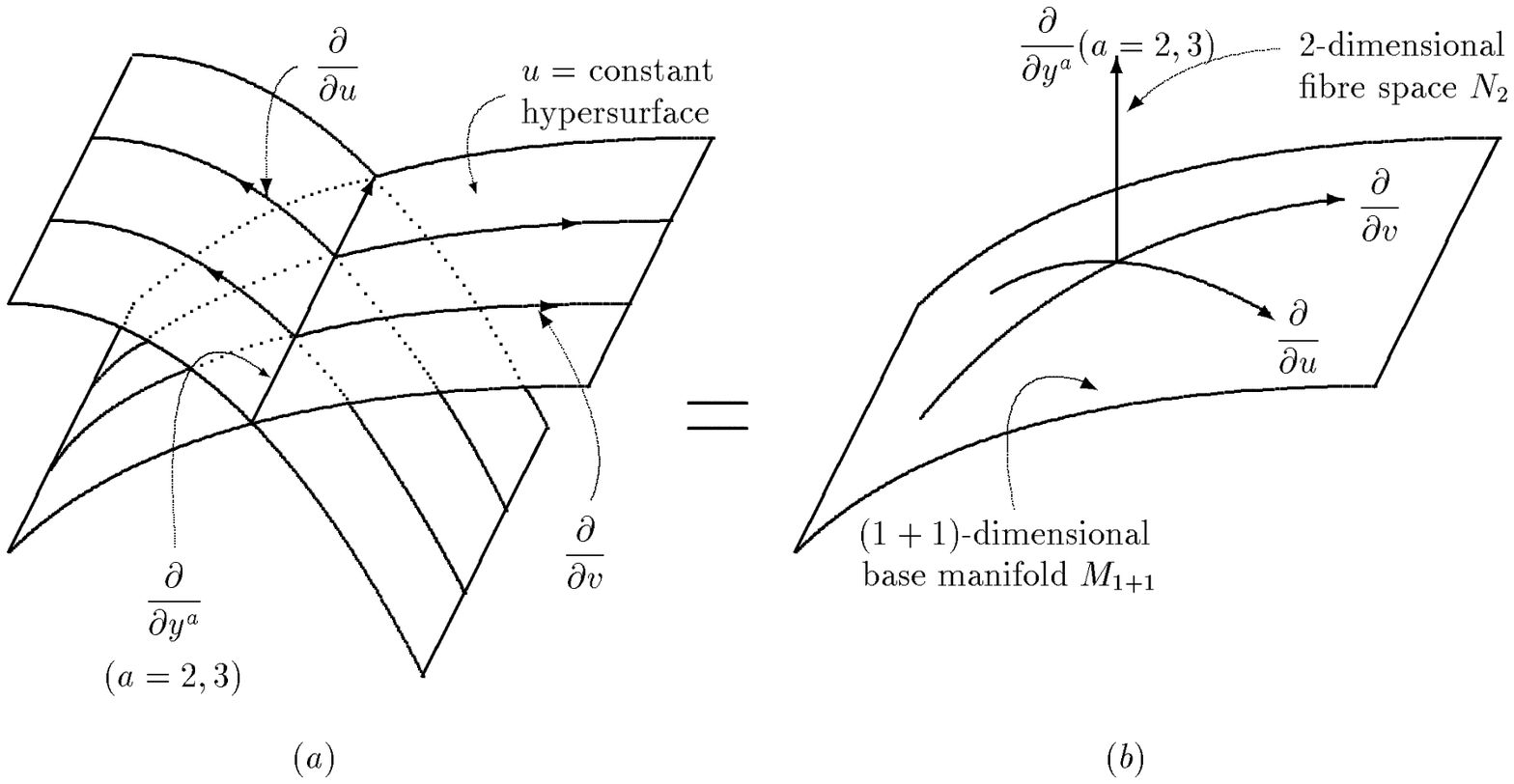}}
\vspace{-11cm}
\caption{The relation between the kinematics of the
null hypersurface formalism and that of the Kaluza-Klein formalism
in the (2,2)-splitting of spacetime is shown.\\
(a) The null hypersurface formalism:
    a null hypersurface labelled by $u={\rm constant}$ is 
    generated by
    a null vector field $\partial/\partial v$ and two transverse
    vector fields $\partial/\partial y^{a}$.\\
(b) The Kaluza-Klein formalism: this shows the KK picture 
    of the same geometry
    from the fibre bundle point of view in the (2,2)-splitting
    of spacetime. Here the (1+1)-dimensional base manifold $M_{1+1}$ 
    is generated by the vector fields $\partial/\partial u$ and 
    $\partial/\partial v$, and the 2-dimensional fibre space 
    $N_{2}$ is generated by the vector fields 
    $\partial/\partial y^{a}$.}
\label{fig1}
\end{figure}

In the Newman-Unti coordinate, these null tetrads 
can be explicitly constructed as follows. 
The first null tetrad $l$ is given by
\begin{equation}
l=l^{A}{\partial \over \partial x^{A}}
={\partial \over \partial v},              \label{el}
\end{equation}
which follows from the equation (\ref{els}).
The second null tetrad $n$ is chosen such that
the following condition
\begin{equation}
l^{A}n_{A}=-1                  \label{condition}
\end{equation}
holds. The most general form of $n$ compatible with this
condition can be written as
\begin{equation}
n=n^{A}{\partial \over \partial x^{A}}
  ={\partial \over \partial u} + U{\partial \over \partial v}
  + X^{a}{\partial \over \partial y^{a}},   \label{en}
\end{equation}
as can be easily confirmed by inspection.
The complex null tetrads $m$ and $\bar{m}$, which satisfy 
the defining relation (\ref{tet}), with $l$ and $n$ given 
by (\ref{el}) and (\ref{en}), respectively, become
\begin{eqnarray}
& &m=m^{A}{\partial \over \partial x^{A}}
  = \omega{\partial \over \partial v}
  +\xi^{a}{\partial \over \partial y^{a}},     \label{em}\\
& &\bar{m}=\bar{m}^{A}{\partial \over \partial x^{A}}
  =\bar{\omega}{\partial \over \partial v}
  +\bar{\xi}^{a}{\partial \over \partial y^{a}}, \label{embar}
\end{eqnarray}
where $\omega$ and $\xi^{a}$ are complex 
quantities\footnote{
Here we may choose the orientation of 
the complex null tetrad $m$ such that $\omega=0$, 
since ${\partial / \partial v}$ has a zero norm.}.
Putting together (\ref{el}), (\ref{en}), (\ref{em}) and (\ref{embar}) 
into (\ref{conditiona}), we find that 
\begin{equation}
\begin{array}{ll}
g^{++}=g^{+a}=0, & g^{+-}=-1, \nonumber\\
g^{--}=-2U + 2 \omega \bar{\omega}, &
g^{-a}=-X^{a} +  \omega \bar{\xi}^{a} 
+\bar{\omega}\xi^{a},                 \nonumber\\
g^{ab}=2 \xi^{{(}a}\bar{\xi}^{b{)}}. &  \label{resz}
\end{array}
\end{equation}
If we compare these metric coefficients with the ones 
given in (\ref{fixed}) in section \ref{s3},
we find that the KK variables can be written in terms of
the null tetrads as 
\begin{eqnarray}
& &2h=-2U + 2\omega \bar{\omega},        \label{nicb}   \\
& &A_{+}^{\ a}=-X^{a} + \omega \bar{\xi}^{a}
   +\bar{\omega} {\xi}^{a},               \label{nicc}   \\
& &\phi^{a b}=2 \xi^{{(}a}\bar{\xi}^{b{)}}.  \label{nice}
\end{eqnarray}
Equivalently, the null tetrads can be written as
\begin{eqnarray}
& &l={\partial \over \partial v},                \\
& &n={\partial \over \partial u} 
  + \Big( -h + \omega\bar{\omega} \Big) {\partial \over \partial v}
  + \Big( -A_{+}^{\ a}+\omega\bar{\xi}^{a}+ \bar{\omega}\xi^{a} \Big)
  {\partial \over \partial y^{a}},               \\
& &m=\omega{\partial \over \partial v}
  +\xi^{a}{\partial \over \partial y^{a}}, 
\, \ {\rm and} \, \ {\rm c.c.}.
\end{eqnarray}
\end{section}

\begin{section}{Classification of the Field Equations}
\label{s6}

It will be instructive to find how the KK field equations 
(\ref{cpm}), $\cdots$, $(\ref{eqc})$ in section \ref{s4} 
are related to the Einstein's equations 
in the standard null hypersurface formalism.
Let us recall that, 
in the standard null hypersurface formalism,
the Einstein's equations are classified into six main equations,
one trivial equation, and three supplementary equations.
Let $G^{AB}$ be the Einstein tensor. 
Then the Einstein's equations are classified as
\begin{eqnarray}
&(A)& \hspace{.5cm}{\rm main} \ {\rm equations}: 
  G^{AB}l_{B}=0, 
\hspace{.5cm} G^{AB}m_{A}m_{B}=0    \label{clast} \\
&(B)&  \hspace{.5cm} {\rm trivial} \ {\rm equation}: 
  G^{AB}m_{A}\bar{m}_{B}=0          \label{class}  \\
&(C)&  \hspace{.5cm}{\rm supplementary} \ {\rm equations}:
  G^{AB}n_{A}m_{B}=0, \hspace{.5cm}
  G^{AB}n_{A}n_{B}=0.                      \label{clasu} 
\end{eqnarray}
Moreover, the Bianchi identity $G^{A B}_{\ \ \ ; B}=0$
can be used to prove the following lemma\cite{sach}:
Suppose that the main equations hold everywhere; 
then (i) the trivial equation holds
everywhere, and (ii) the supplementary equations hold 
everywhere if they hold at some $v={\rm constant}$ hypersurface.

In order to see what these classified equations are like 
in the Newman-Unti coordinate, we have to find 
the covariant null tetrads $(l_{A}, \ n_{A},\ m_{A}, \ \bar{m}_{A})$
in the Newman-Unti coordinate first, and then form 
the above combinations using them. 
Notice that $l_{A}$ is already given by the equation (\ref{els})
in section \ref{s5}.
To find $m_{A}$, we notice that 
\begin{eqnarray}
m_{A}&=&m^{B}g_{AB}                 \nonumber\\
     &=&\omega g_{A-} + \xi^{a}g_{A a},     \label{coem}
\end{eqnarray}
since $m^{+}=0$, $m^{-}=\omega$, and $m^{a}=\xi^{a}$.
Thus, using $g_{AB}$ in (\ref{resa}) in section \ref{s3}, we find that 
the coefficients of the complex null tetrad $m$ are given by
\begin{eqnarray}
& &m_{+}= - \omega + \xi_{a}A_{+}^{\ a}, \\
& &m_{-}=0,                              \\
& &m_{a}=\xi_{a},         
\end{eqnarray}
where $\xi_{a}=\phi_{a b}\xi^{b}$.
Similarly, the equation 
\begin{equation}
n_{A}=n^{B}g_{AB}
\end{equation}
gives
\begin{eqnarray}
n_{+}&=&-2h - U  +\Big( 
  \omega\bar{\xi}_{a} 
+ \bar{\omega}\xi_{a} \Big)A_{+}^{\ a},   \label{tttr}\\
n_{-}&=&-1,                              \label{tttq}\\
n_{a}&=&\omega\bar{\xi}_{a} + \bar{\omega}\xi_{a}, \label{tttb}
\end{eqnarray}
where we used the equation (\ref{en}) and (\ref{nicc})
in section \ref{s5}.

Let us now write the main equations, the trivial equation, and 
the supplementary equations in the Newman-Unti coordinate.
$G^{AB}l_{B}$ can be written as
\begin{equation}
G^{AB}l_{B}= -G^{AB}\delta _{B}^{ \ +} =-G^{+A}.   \label{omit}
\end{equation}
Therefore we have
\begin{equation}
G^{AB}l_{B}=0
\Longleftrightarrow
\Big\{ G^{++}=0, \hspace{.5cm}  G^{+-}=0, 
       \hspace{.5cm}  G^{+a}=0  \Big\}.   \label{split}
\end{equation}
Also $G^{AB}m_{A}m_{B}$ becomes
\begin{eqnarray}
G^{AB}m_{A}m_{B}
&=&G^{++}m_{+}m_{+}+2G^{+a}m_{+}m_{a} 
+ G^{ab}m_{a}m_{b} \nonumber\\
&=&G^{ab}m_{a}m_{b}               \nonumber\\
&=&G^{ab}\xi_{a}\xi_{b},                   \label{emem}
\end{eqnarray}
where we used the equation (\ref{split}). 
Thus the main equations (\ref{clast}) are
equivalent to the following set of equations
\begin{equation}
\left\{
\begin{array}{l}
         G^{AB}l_{B}=0 \nonumber\\
         G^{AB}m_{A}m_{B}=0 
\end{array}
\right\}
\Longleftrightarrow
\left\{ \begin{array}{l}
        G^{++}=0, \hspace{.5cm} G^{+-}=0, \hspace{.5cm}
        G^{+a}=0 \nonumber\\
        G^{ab}\xi_{a}\xi_{b}=0
           \end{array}
\right\}.                        \label{maineq}
\end{equation}
Notice that $G^{AB}m_{A}\bar{m}_{B}$ becomes,
\begin{eqnarray}
G^{AB}m_{A}\bar{m}_{B}
&=&G^{++}m_{+}\bar{m}_{+} + G^{+a}m_{+}\bar{m}_{a}
  +G^{a+}m_{a}\bar{m}_{+} + G^{ab}m_{a}\bar{m}_{b} \nonumber\\
&=&G^{ab}m_{{(}a}\bar{m}_{b{)}}       \nonumber\\
&=&G^{ab}\xi_{{(}a}\bar{\xi}_{b{)}},   \label{tri}
\end{eqnarray}
using the main equations. 
Therefore we have the following correspondence
\begin{equation}
G^{AB}m_{A}\bar{m}_{B}=0 \Longleftrightarrow 
G^{ab}\xi_{{(}a}\bar{\xi}_{b{)}}=0,      \label{triz}
\end{equation}
assuming that the main equations hold.
Let us consider the supplementary equations now.
Using the main equations (\ref{maineq}), 
$G^{AB}n_{A}m_{B}$ becomes 
\begin{eqnarray}
G^{AB}n_{A}m_{B}&=&G^{++}n_{+}m_{+}+G^{+a}n_{+}m_{a}
+G^{-+}n_{-}m_{+}+G^{-a}n_{-}m_{a} 
+G^{a+}n_{a}m_{+}           \nonumber\\
& &+G^{ab}n_{a}m_{b} \nonumber\\
&=&G^{-a}n_{-}m_{a}+ G^{ab}n_{a}m_{b}      \nonumber\\
&=&-G^{-a}\xi_{a} 
+ \bar{\omega}G^{ab}\xi_{a}\xi_{b}
+ \omega G^{ab}\bar{\xi}_{{(}a}\xi_{b{)}} \nonumber\\
&=&-G^{-a}\xi_{a},                 \label{tria}
\end{eqnarray}
where in the last line we used the equations
\begin{equation}
G^{ab}\xi_{a}\xi_{b}=0
= G^{ab}\bar{\xi}_{{(}a}\xi_{b{)}},      \label{both}    
\end{equation}
which follow from the equations  (\ref{maineq}) and (\ref{triz}).
Thus, provided that the main equations and the trivial 
equation hold, we find that 
\begin{equation}
G^{AB}n_{A}m_{B}=0=-G^{-a}\xi_{a}.          \label{gooda}
\end{equation}
This equation, together with its complex conjugate
\begin{equation}
G^{-a}\bar{\xi}_{a}=0,
\end{equation}
yields 
\begin{equation}
2G^{-a}\xi_{{(}a}\bar{\xi}_{b{)}}=G^{-a}\phi_{a b}=0.
\end{equation}
This shows that the following sets of equations are equivalent,
\begin{equation}
G^{AB}n_{A}m_{B}=0 
\Longleftrightarrow G^{-a}=0,                \label{ccc}
\end{equation}
provided the main equations and the trivial equation hold.
Finally let us notice that $G^{AB}n_{A}n_{B}$ becomes
\begin{eqnarray}
G^{AB}n_{A}n_{B}
&=&G^{++}n_{+}n_{+} + G^{--}n_{-}n_{-} +G^{ab}n_{a}n_{b} 
+2G^{+-}n_{+}n_{-}         
+ 2 G^{+a}n_{+}n_{a}            \nonumber\\
& &+ 2G^{-a}n_{-}n_{a}           \nonumber\\
&=&G^{--}n_{-}n_{-}  +G^{ab}n_{a}n_{b} 
 + 2G^{-a}n_{-}n_{a}             \nonumber\\
&=&G^{--}+G^{ab}\Big( \omega\bar{\xi}_{a}+\bar{\omega}\xi_{a}\Big)
\Big( \omega\bar{\xi}_{b}
+\bar{\omega}\xi_{b}
\Big)
-2G^{-a}\Big( \omega \bar{\xi}_{a} + \bar{\omega}\xi_{a}
        \Big)                      \nonumber\\
&=&G^{--},           \label{final}
\end{eqnarray}
where we again used the main equations and the trivial equation.
From (\ref{ccc}) and (\ref{final}), we find that the followings
are equivalent:
\begin{equation}
\left\{
\begin{array}{l}
G^{AB}n_{A}m_{B}=0 \nonumber\\
G^{AB}n_{A}n_{B}=0
\end{array}
\right\}
\Longleftrightarrow
\left\{ \begin{array}{l}
        G^{-a}=0     \nonumber\\ 
        G^{--}=0
        \end{array}
\right\}.                        \label{finalz}      
\end{equation}
Putting these results together, 
the Einstein's equations in the standard null hypersurface 
formalism become, in the Newman-Unti coordinate, as follows:
\begin{eqnarray}
&(A)& \hspace{.1cm} {\rm main} \ {\rm equations}:
\left\{
\begin{array}{l}
G^{A B}l_{B}=0 \nonumber\\
G^{A B}m_{A}m_{B}=0 
\end{array} \right\}
\Longleftrightarrow
\left\{
\begin{array}{l}
G^{++}=0, \, G^{+-}=0, \, G^{+a}=0    \nonumber\\
G^{ab}\xi_{a}\xi_{b}=0
\end{array} \right\},          \label{xxx}\\
&(B)& \hspace{.1cm} {\rm trivial} \ {\rm equation}:
G^{A B}m_{A}\bar{m}_{B}=0
\Longleftrightarrow 
G^{ab}\xi_{{(}a}\bar{\xi}_{b{)}}=0,    \label{xxy}\\
&(C)&\hspace{.1cm} {\rm supplementary} \ {\rm equations}:
\left\{
\begin{array}{l}
G^{A B}n_{A}m_{B}=0 \nonumber\\
G^{A B}n_{A}n_{B}=0
\end{array}\right\}
\Longleftrightarrow
\left\{
\begin{array}{l}
G^{-a}=0\nonumber\\
G^{--}=0
\end{array}\right\}.        \label{xxz}
\end{eqnarray}

Having expressed the main equations, trivial equation, and
supplementary equations in the Newman-Unti coordinate, 
we now wish to group the KK field equations 
(\ref{cpm}), $\cdots$, (\ref{eqc})
in accordance with the above classification. 
Let $S$ be the Einstein-Hilbert action.
The variation of $S$ with respect to the 
metric\footnote{Here the notations are as before, i.e.
$A=(\mu, a); \mu=(0,1) \ {\rm or} \ (+,-), a=(2,3)$.} 
$g_{A B}$ is given by
\begin{eqnarray}
\delta S &=&\int G^{AB}\delta g_{A B} 
+ {\rm surface} \ {\rm terms}       \nonumber\\        
&=&\int G^{\mu\nu}\delta g_{\mu\nu}  
+\int 2 G^{\mu a}\delta g_{\mu a}
+\int G^{a b}\delta g_{a b}
+{\rm surface} \ {\rm terms},            \label{obs}
\end{eqnarray}
where the integration is over 4-dimensional spacetime.
In terms of the KK variables (\ref{gen}) in section \ref{s2}, 
the metric variations can be written as
\begin{eqnarray}
& &\delta g_{\mu\nu}=\delta \gamma_{\mu\nu}
+A_{\mu}^{\ a}A_{\nu}^{\ b}\delta \phi_{a b}
+\phi_{a b}A_{\mu}^{\ a}\delta A_{\nu}^{\ b}
+\phi_{a b}A_{\nu}^{\ b}\delta A_{\mu}^{\ a}, \label{relx}\\
& &\delta g_{\mu a}=A_{\mu}^{\ b}\delta \phi_{a b}
+\phi_{a b}\delta A_{\mu}^{\ b},               \label{rely}\\
& &\delta g_{a b}=\delta \phi_{a b}.          \label{rela}
\end{eqnarray}
Therefore $\delta S$ becomes
\begin{eqnarray}
\delta S &=&\int G^{\mu\nu}\delta \gamma_{\mu\nu}
  +\int 2\Big ( G^{\mu\nu}\phi_{a b}A_{\nu}^{\ b}
  +G^{\mu b}\phi_{a b} \Big) \delta A_{\mu}^{\ a}  \nonumber\\
& &+\int \Big( G^{\mu\nu} A_{\mu}^{\ a} A_{\nu}^{\ b}
  +2G^{\mu a}A_{\mu}^{\ b} + G^{a b} \Big)
  \delta \phi_{a b}+{\rm surface} \ {\rm terms}.  \label{relb}
\end{eqnarray}
If we equate the coefficients of each variation in (\ref{relb})
to zero, we could obtain the Einstein's equations written 
in terms of the general KK variables without any gauge-fixing.
If we use the gauge condition (\ref{res})
in section \ref{s3} at the end of the above variations, 
then we find that\footnote{Here the subscript ${}_{\rm p}$ 
in $G_{\rm p}^{AB}$ stands for the Einstein tensor $G^{AB}$ 
evaluated in the Polyakov gauge (\ref{res}) in section \ref{s3}.}
\begin{eqnarray}
\delta S &=& \int \Big\{ 
G_{\rm p}^{++}\delta \gamma_{++} 
+2 G_{\rm p}^{+-}\delta \gamma_{+-} 
+G_{\rm p}^{--}\delta \gamma_{--}    
+2\Big( G_{\rm p}^{++}\phi_{a b} A_{+}^{\ b} 
         + G_{\rm p}^{+-}\phi_{a b} A_{-}^{\ b} 
+  G_{\rm p}^{+b}\phi_{a b}\Big)
\delta A_{+}^{\ a}            \nonumber\\
& &+2\Big( G_{\rm p}^{-+}\phi_{a b}A_{+}^{\ b} 
      + G_{\rm p}^{--}\phi_{a b}A_{-}^{\ b} 
+ G_{\rm p}^{-b}\phi_{a b}\Big)
\delta A_{-}^{\ a}            \nonumber\\
& &+\Big( G_{\rm p}^{++} A_{+}^{\ a} A_{+}^{\ b} 
+2G_{\rm p}^{+-} A_{+}^{\ a} A_{-}^{\ b}
+G_{\rm p}^{--} A_{-}^{\ a} A_{-}^{\ b}
+2G_{\rm p}^{+a} A_{+}^{\ b} 
+2G_{\rm p}^{-a} A_{-}^{\ b} 
+G_{\rm p}^{ab}\Big)\delta \phi_{a b} \Big\} \nonumber\\
& &+{\rm surface} \ {\rm terms}.      \label{great}
\end{eqnarray}
The equations resulting from these variations
will be precisely the KK field equations,
i.e. the equations (\ref{cpm}), $\cdots$, (\ref{eqc}) 
in section \ref{s4}. 
Notice that the equations of motion of $\phi_{a b}$, 
\begin{equation}
G_{\rm p}^{++} A_{+}^{\ a} A_{+}^{\ b} 
+2G_{\rm p}^{+-} A_{+}^{\ a} A_{-}^{\ b}
+G_{\rm p}^{--} A_{-}^{\ a} A_{-}^{\ b}
+2G_{\rm p}^{+a} A_{+}^{\ b} 
+2G_{\rm p}^{-a} A_{-}^{\ b} 
+G_{\rm p}^{ab}=0, \label{facto}
\end{equation}
can be split into three equations by taking 
trace with $\phi_{a b}$, 
and by multiplying with $\xi_{a}\xi_{b}$ and its complex conjugate.
The resultant equations are 
\begin{eqnarray}
&(d)&\hspace{.5cm} 
G_{\rm p}^{++}\phi_{a b} A_{+}^{\ a} A_{+}^{\ b} 
+2G_{\rm p}^{+-}\phi_{a b} A_{+}^{\ a} A_{-}^{\ b}
+G_{\rm p}^{--}\phi_{a b} A_{-}^{\ a} A_{-}^{\ b}
+2G_{\rm p}^{+a}\phi_{a b} A_{+}^{\ b}     \nonumber\\
& &+2G_{\rm p}^{-a}\phi_{a b} A_{-}^{\ b} 
+G_{\rm p}^{ab}\phi_{a b}=0,     \label{factoa}\\
&(e)&\hspace{.5cm} G_{\rm p}^{++} (\xi_{a}A_{+}^{\ a})^{2} 
+2G_{\rm p}^{+-} (\xi_{a}A_{+}^{\ a})(\xi_{b} A_{-}^{\ b})
+G_{\rm p}^{--}(\xi_{a} A_{-}^{\ a})^{2}
+2G_{\rm p}^{+a} \xi_{a} \xi_{b}A_{+}^{\ b} \nonumber\\
& &+2G_{\rm p}^{-a}\xi_{a}\xi_{b} A_{-}^{\ b} 
+G_{\rm p}^{ab}\xi_{a}\xi_{b}=0, 
\, \ {\rm and} \, \ {\rm c.c.},     \label{separ}
\end{eqnarray}
which correspond to the equations (\ref{eqb}) and (\ref{prop}) 
in section \ref{s4} obtained by the variations with respect to
$\sigma$ and $\rho_{a b}\ ({\rm det}\rho_{a b}=1)$.
Thus, together with the equations (\ref{factoa}) and (\ref{separ}), 
the complete set of the 
KK field equations in terms of the Einstein tensor becomes,
in the Polyakov gauge, 
\begin{eqnarray}
&(a)&\hspace{.5cm} G_{\rm p}^{++}=0,             \label{ei}\\
&(b)& \hspace{.5cm}
      G_{\rm p}^{++}\phi_{a b} A_{+}^{\ b} 
     +G_{\rm p}^{+-}\phi_{a b} A_{-}^{\ b} 
     +G_{\rm p}^{+b}\phi_{a b}=0,           \label{bi}\\
&(c)&\hspace{.5cm} 
G_{\rm p}^{-+}\phi_{a b}A_{+}^{\ b} 
      + G_{\rm p}^{--}\phi_{a b}A_{-}^{\ b} 
+ G_{\rm p}^{-b}\phi_{a b}=0,           \label{ci}\\
&(f)&\hspace{.5cm} G_{\rm p}^{+-}=0,                   \label{ef}\\
&(g)&\hspace{.5cm} G_{\rm p}^{--}=0.                \label{gi}
\end{eqnarray}
Notice that the right hand side of the classification 
(\ref{xxx}) requires the main equations be valid throughout.
Thus, if we assume that
$G_{\rm p}^{++}=0$, $G_{\rm p}^{+-}=0$, $G_{\rm p}^{+a}=0$, 
and $A_{-}^{\ a}=0$, the equations (\ref{factoa}), $\cdots$, 
(\ref{gi}) become,
\begin{eqnarray}
&(a)& \hspace{.5cm}G_{\rm p}^{++}=0,             \label{eia}\\
&(b)& \hspace{.5cm}G_{\rm p}^{+a}=0,             \label{bia}\\
&(c)&\hspace{.5cm} G_{\rm p}^{-a}=0,             \label{cia}\\
&(d)&\hspace{.5cm} 
     G_{\rm p}^{ab}\xi_{{(}a}\bar{\xi}_{b{)}} =0, \label{dia}\\
&(e)&\hspace{.5cm} G_{\rm p}^{ab}\xi_{{(}a}\xi_{b{)}}=0,
\,\ {\rm and} \, \ {\rm c.c.},         \label{eea}\\
&(f)&\hspace{.5cm} G_{\rm p}^{+-}=0,                     \label{efa}\\
&(g)&\hspace{.5cm} G_{\rm p}^{--}=0.               \label{gia}
\end{eqnarray}

It is well-known that 
the four equations $G^{-A}=0$, when $x^{-}=v$ is 
viewed as the time coordinate, are the four constraint equations 
of the vacuum general relativity. These are the equations 
(\ref{cia}), (\ref{efa}), and (\ref{gia}), which correspond 
in the KK formalism to 
the constraint equations (\ref{eqa}), (\ref{eqd}), and (\ref{eqc}) 
in section \ref{s4}, since only the first $v$-time 
derivatives appear in these equations.

\end{section}

\begin{section}{Discussions}
\label{s7}

In this paper, we presented the KK formalism of general
relativity of 4-dimensional spacetimes, 
viewing the spacetime as a fibred manifold, namely,
a local product of the (1+1)-dimensional base manifold and 
the 2-dimensional fibre space. 
Within this framework, we made a decomposition
of a given 4-dimensional spacetime metric into sets of fields
which transform as a tensor field, gauge fields, and
scalar fields, with respect to the group of diffeomorphisms 
of the 2-dimensional fibre space. 
Moreover, using the Polyakov gauge,
we arrived at an amazingly simple
expression of the Einstein-Hilbert action, i.e. 
a Yang-Mills type gauge theory action coupled to 
a non-linear sigma field and a scalar field.
Then we presented the Einstein's equations in terms of 
these fields.
 
This KK approach has several advantages, as we already mentioned.
In connection with issues of quantum gravity, this approach
has the following aspects which deserve further remarks.
First, the problem of solving the Einstein's constraint 
equations is known to be one of the most important tasks 
in quantum general relativity. 
Remarkably, in the (2,2)-KK approach, the diffeomorphisms 
of the 2-dimensional fibre space turns out to be a
local gauge symmetry, exactly as in Yang-Mills theory. 
Therefore the two constraint
equations associated with these diffeomorphisms
can be automatically solved, 
by using the diff$N_{2}$-invariant quantities. 
However, there are two additional 
constraint equations which require further studies\cite{solo}
in order to fully take care of the four Einstein's 
constraint equations.

Second, the Polyakov gauge is shown to be equivalent 
to the Newman-Unti coordinate, modulo the diff$N_{2}$ 
transformations. It is well-known that, 
in the null hypersurface formalism 
in the Newman-Unti coordinate, the physical gravitational
degrees of freedom can be easily isolated, whereas in the standard 
(3+1)-formalism or in the ADM formalism such an isolation 
can be done only in the linearized theory. Hence the KK formalism 
in the Polyakov gauge shares this salient feature of
the null hypersurface formalism:
it is the non-linear sigma field
$\rho_{ab}$ with a unit determinant that is precisely the physical
gravitational degrees of freedom, which are also known as
the conformal 2-metric of the transverse 2-surface 
in the null hypersurface formalism.

Third, it is also of interest to find how the lightcones 
fluctuate in quantum gravity. 
In the formalism it is quite easy 
to address this issue. Recall that a lightcone is 
a collection of null vector fields
originating from a given spacetime point. The relevant null 
vector field is, as given in section \ref{s5} 
(with $\omega=0$),
\begin{equation}
n={\partial\over \partial u}-h{\partial\over \partial v}
  -A_{+}^{\ a}{\partial\over \partial y^{a}},
\end{equation}
which clearly has a zero norm.
This vector field is controlled by the fields
$h$ and $A_{+}^{\ a}$. Therefore, by quantizing these fields, we
can quantize the lightcone itself
without referring to any background spacetime geometry.

Fourth, it should be mentioned that the Lie derivative appears 
naturally
in this formalism, via the {\it minimal} couplings to the 
diff$N_{2}$-valued gauge fields. In the standard 
(3+1)-formalism, the natural derivative operator is the 
metric-compatible covariant derivative, which requires the metric 
be non-degenerate. The Lie derivative, on the other hand, can be
defined even when the metric is degenerate. Therefore,
the KK formalism, based on the notion of the Lie derivative,
should be extendable to spacetimes where the metric is degenerate, 
which would be hard to describe 
in conventional approaches.

One may also attack more ``practical'' problems in classical 
general relativity. For instance, we expect the KK formalism to be
useful for problems such as finding exact
solutions of the Einstein's equations\cite{sssol},
in view of the fact that 
the null hypersurface formalism, particularly the Newman-Penrose
version, has been proven extremely useful for such problems.
To those problems concerned with asymptotically flat radiative
spacetimes\cite{solo} or gravitational waves in general, 
the new formalism may be also applicable. 
A related but probably more challenging
problem is to reinterpret {\it all} the known exact solutions
of the Einstein's equations from the perspective of 
the Yang-Mills type gauge theories coupled to the non-linear sigma
field and a scalar field on the (1+1)-dimensional base manifold.
This seems very interesting, for there are a number of exact
solutions of the Einstein's equations which do not permit a sensible
physical interpretation from the 4-dimensional spacetime 
perspective\cite{exact}.

Finally, we would like to remark on the question of finiteness
of quantum gravity. Since the question whether a theory 
is renormalizable or not depends critically on 
what spacetime dimensions
the theory is defined on, this (1+1)-dimensional description of
general relativity of 4-dimensions may lead to a happy conclusion 
that quantum gravity is, after all, a {\it finite} theory, 
at least when 
viewed from this (1+1)-dimensional field theory perspective.
\end{section}

\vskip 1.5cm
\noindent
\centerline{\bf Acknowledgments}\\

This work is supported in part by the Basic Science
Research Institute Program, Ministry of Education,
1995 (BSRI-95-2418), and by non-directed research fund, 
Korea Research Foundation, 1995 (04-D-0073). 
It is also supported by Korea Science and Engineering 
Foundation (95-0702-04-01-3), and by the SRC program 
of SNU-CTP of KOSEF. Parts of this work were done
while the author was a post-doctoral fellow at FAMAF,
the Universidad Nacional de C\'{o}rdoba, Argentina.
\bigskip\goodbreak

\appendix
\section{Diff$N_{2}$ Transformations}
\label{a1}

In this Appendix we shall first derive the finite transformation 
properties (\ref{phi}), (\ref{newtran}), and (\ref{gam}) 
and the corresponding infinitesimal transformation 
properties (\ref{var2}), (\ref{var1}), and (\ref{var3}), 
and then show that $D_{\mu}\phi_{ab}$ and $F_{\mu\nu}^{\ \ a}$
transform covariantly under the diff$N_{2}$ transformations,
as given in (\ref{fifc}) and (\ref{fifa}) in section \ref{s2}. 
The spacetime metric at the point
$(x^{\mu},y^{a})$ is given by (\ref{gen}) in section \ref{s2},
\begin{eqnarray}
ds^{2}&=&\gamma_{\mu\nu}dx^{\mu}dx^{\nu}
+ \phi_{a b}\Big( dy^{a} + A_{\mu}^{\ a}dx^{\mu} \Big)
            \Big( dy^{b} + A_{\nu}^{\ b}dx^{\nu}  \Big)\nonumber\\
&=&\Big( \gamma_{\mu\nu}
   + \phi_{a b}A_{\mu}^{\ a}A_{\nu}^{\ b}\Big)dx^{\mu}dx^{\nu}
   + 2\phi_{a b}A_{\mu}^{\ a}dx^{\mu}dy^{b} 
+ \phi_{a b}dy^{a}dy^{b}.       \label{agen}
\end{eqnarray}
Let us consider the following arbitrary transformation of 
the $y^{a}$-coordinate at each point of the base manifold $x^{\mu}$, 
while keeping $x^{\mu}$ constant, 
\begin{eqnarray}
& &y^{' a}=y^{' a} (x,y),               \nonumber\\
& &x^{' \mu}=x^{\mu},                   \label{anew}
\end{eqnarray}
where $y^{' a} (x,y)$ is an arbitrary function of 
$x^{\mu}$ and $y^{a}$. Then we have
\begin{eqnarray}
& &dy^{'a}=\Big( {\partial y^{'a}\over \partial y^{c}}\Big) dy^{c}
+\Big( 
{\partial y^{'a}\over \partial x^{\mu}}\Big) dx^{\mu},  \nonumber\\
& &dx^{' \mu}=dx^{\mu},
\label{newa}
\end{eqnarray}
or equivalently,
\begin{eqnarray}
& &dy^{a}={\partial y^{a}\over \partial y^{'c}}
\Big\{ dy^{'c} - \Big( {\partial y^{'c}\over \partial x^{\mu}}\Big)
  dx^{'\mu} \Big\},                 \nonumber\\
& &dx^{\mu}=dx^{' \mu}.            \label{newb}
\end{eqnarray}
Thus we can express the metric at the point $(x^{\mu},y^{a})$ 
in terms of the differentials of the new coordinates 
$dx^{'\mu}$ and $dy^{'a}$. The term proportional to
$dx^{\mu}dy^{a}$ in (\ref{agen}) becomes
\begin{eqnarray}
& &2\phi_{a b}(x,y)A_{\mu}^{\ a}(x,y)dx^{\mu}dy^{b} \nonumber\\
&=&2\phi_{a b}(x,y)A_{\mu}^{\ a}(x,y)
 dx^{'\mu}
 \Big( {\partial y^{b}\over \partial y^{'c}} \Big)
 \Big\{ dy^{'c} -
\Big( {\partial y^{'c} \over \partial x^{\nu}}\Big) dx^{'\nu} 
              \Big\}                      \nonumber\\
&=&2\Big( {\partial y^{b}\over \partial y^{'c}}\Big)
\phi_{a b}(x,y)\Big( {\partial y^{a}\over \partial y^{'d}}\Big)
\Big(  {\partial y^{'d}\over \partial y^{e}}\Big)
A_{\mu}^{\ e}(x,y)dx^{'\mu}
\Big\{ dy^{'c} -
\Big( {\partial y^{'c} \over \partial x^{\nu}} \Big)
dx^{'\nu} \Big\}                       \nonumber\\
&=&2\Big( {\partial y^{a}\over \partial y^{'c}} \Big)
 \Big( {\partial y^{b}\over \partial y^{'d}} \Big) \phi_{a b}(x,y)
\Big(  {\partial y^{'d}\over \partial y^{e}} \Big)
A_{\mu}^{\ e}(x,y)dx^{'\mu}
\Big\{ dy^{'c} -
\Big( {\partial y^{'c} \over \partial x^{\nu}}\Big)
dx^{'\nu} \Big\},                   \label{newd}
\end{eqnarray}
where the $(x^{\mu},y^{a})$-dependence of each field 
is shown explicitly, and the identity
\begin{equation}
\Big( {\partial y^{a}\over \partial y^{'d}} \Big)
\Big( {\partial y^{'d}\over \partial y^{e}} \Big)
=\delta_{e}^{\ a}
\end{equation}
was used. 
The term containing $dy^{a}dy^{b}$ becomes
\begin{eqnarray}
& &\phi_{a b}(x,y)dy^{a}dy^{b}            \nonumber\\
&=&\phi_{a b}(x,y)
\Big( {\partial y^{a}\over \partial y^{'c}} \Big)
\Big(   {\partial y^{b}\over \partial y^{'d}} \Big)
  \Big\{  dy^{'c}
 -\Big( {\partial y^{'c} \over \partial x^{\mu}}\Big) dx^{' \mu} 
  \Big\}
  \Big\{ dy^{'d}
 - \Big( {\partial y^{'d} \over \partial x^{\nu}}\Big)
   dx^{' \nu} \Big\}  \nonumber\\
&=&\Big(   {\partial y^{a}\over \partial y^{'c}}\Big)
\Big( {\partial y^{b}\over \partial y^{'d}}\Big) \phi_{a b}(x,y)
  \Big\{
  dy^{'c} dy^{'d}
  -2\Big( {\partial y^{'d} \over \partial x^{\mu}}\Big) 
  dy^{'c}dx^{' \mu}
 +\Big( {\partial y^{'c}\over \partial x^{\mu}}\Big)
 \Big( {\partial y^{'d}\over \partial x^{\nu}}\Big)
  dx^{'\mu}dx^{'\nu}\Big\}.            \label{newc}
\end{eqnarray}
Therefore the line element (\ref{agen}) becomes
\begin{eqnarray}
ds^{2}&=&\Big\{ \gamma_{\mu\nu}(x,y)
   +\phi_{a b}(x,y)A_{\mu}^{\ a}(x,y)A_{\nu}^{\ b}(x,y) \Big\}
   dx^{'\mu}dx^{'\nu}                  \nonumber\\
& &+\Big( {\partial y^{a}\over \partial y^{'c}}\Big)
  \Big( {\partial y^{b}\over \partial y^{'d}}\Big)
\phi_{a b}(x,y) \Big\{
  dy^{'c} dy^{'d}
  -2\Big( {\partial y^{'d} \over \partial x^{\mu}}\Big) 
dy^{'c}dx^{'\mu}
 +\Big( {\partial y^{'c}\over \partial x^{\mu}}\Big)
 \Big(  {\partial y^{'d}\over \partial x^{\nu}}\Big)
  dx^{'\mu}dx^{'\nu}\Big\}         \nonumber\\
& &+2\Big( {\partial y^{a}\over \partial y^{'c}}\Big)
  \Big( {\partial y^{b}\over \partial y^{'d}}\Big) \phi_{a b}(x,y)
\Big( {\partial y^{'d}\over \partial y^{e}} \Big)
  A_{\mu}^{\ e}(x,y)  dx^{'\mu}
\Big\{  dy^{'c} -
\Big( {\partial y^{'c} \over \partial x^{\nu}} \Big) 
dx^{'\nu} \Big\}                  \nonumber\\
&=&\gamma_{\mu\nu}(x,y)dx^{'\mu}dx^{'\nu}
  +\Big( {\partial y^{a}\over \partial y^{'c}}\Big)
  \Big( {\partial y^{b}\over \partial y^{'d}}\Big)
  \phi_{a b}(x,y) dy^{'c} dy^{'d} \nonumber\\
& &+2\Big( {\partial y^{a}\over \partial y^{'c}}\Big)
  \Big( {\partial y^{b}\over \partial y^{'d}}\Big)
  \phi_{a b}(x,y)
  \Big\{ 
   \Big( {\partial y^{' d}\over \partial y^{e}}\Big) 
  A_{\mu}^{\ e}(x,y)
  -{\partial y^{' d}\over \partial x^{\mu}} \Big\}
   dx^{'\mu}dy^{'c}\nonumber\\
& &+\phi_{a b}(x,y)  \Big\{
 A_{\mu}^{\ a}(x,y)A_{\nu}^{\ b}(x,y)
- 2\Big( {\partial y^{a}\over \partial y^{'c}}\Big)
  \Big( {\partial y^{b}\over \partial y^{'d}}\Big)
  \Big( {\partial y^{' d}\over \partial y^{e}} \Big) 
  A_{\mu}^{\ e}(x,y)
   \Big( {\partial y^{' c}\over \partial x^{\nu}}  \Big) \nonumber\\
& &+ \Big( {\partial y^{a}\over \partial y^{'c}}\Big)
  \Big( {\partial y^{b}\over \partial y^{'d}}\Big)
 \Big( {\partial y^{' c}\over \partial x^{\mu}} \Big)
 \Big( {\partial y^{' d}\over \partial x^{\nu}} \Big)
\Big\} dx^{'\mu}dx^{'\nu},              \label{large}
\end{eqnarray}
which must be equal to
\begin{equation}
ds^{'2}=\gamma'_{\mu\nu}(x', y')dx^{'\mu}dx^{'\nu}
+ \phi'_{a b}(x', y')\Big\{ dy^{'a} + A_{\mu}^{'\ a}(x', y')
dx^{'\mu} \Big\}
\Big\{ dy^{'b} + A_{\nu}^{'\ b}(x', y')
dx^{'\nu} \Big\},                           \label{newline}
\end{equation}
due to the diffeomorphism invariance.
If we compare terms containing $dy^{'a}dy^{'b}$ 
in (\ref{large}) and (\ref{newline}), we find that
\begin{equation}
\phi'_{a b}(x', y')
=\Big( {\partial y^{c}\over \partial y^{'a}}\Big)
 \Big( {\partial y^{d}\over \partial y^{'b}}\Big)
  \phi_{c d}(x,y),                         \label{coeff1}
\end{equation}
which shows that $\phi_{c d}(x,y)$ is a tensor field
with respect to the diff$N_{2}$ transformations.
Then we may write the line element (\ref{large}) as
\begin {eqnarray}
ds^{2}&=&\gamma_{\mu\nu}(x, y)dx^{'\mu}dx^{'\nu}
 +\phi'_{c d}(x', y')dy^{' c}dy^{' d} 
+2\phi'_{c d}(x', y')\Big\{
 \Big( {\partial y^{' d}\over \partial y^{a}}\Big) A_{\mu}^{\ a}(x, y)
  -{\partial y^{' d}\over \partial x^{\mu}}
   \Big\}
   dx^{'\mu}dy^{'c}               \nonumber\\
& &+\phi'_{c d}(x', y')
   \Big\{
 \Big( {\partial y^{' c}\over \partial y^{a}}\Big) A_{\mu}^{\ a}(x, y)
  -{\partial y^{' c}\over \partial x^{\mu}} \Big\}
  \Big\{
 \Big( {\partial y^{' d}\over \partial y^{b}} \Big) A_{\nu}^{\ b}(x, y)
  -{\partial y^{' d}\over \partial x^{\nu}}\Big\} 
  dx^{'\mu}dx^{'\nu}. 
\end{eqnarray}
From this expression, we find the transformation
properties of $A_{\mu}^{\ a}(x, y)$ and $\gamma_{\mu\nu}(x,y)$,
\begin{eqnarray}
& &A_{\mu}^{'\ a}(x', y')
=\Big( {\partial y^{'a}\over \partial y^{b}}\Big) A_{\mu}^{\ b}(x, y)
-{\partial y^{'a}\over \partial x^{\mu}}(x, y), \label{coeff2}\\
& &\gamma'_{\mu\nu}(x',y')=\gamma_{\mu\nu}(x,y). \label{coeff3}
\end{eqnarray}
The equations (\ref{coeff1}), (\ref{coeff2}), and
(\ref{coeff3}) are the equations (\ref{phi}),
(\ref{newtran}), and (\ref{gam}) in section \ref{s2}.

The infinitesimal versions of the above finite 
transformations can be
also found by considering the following transformations 
\begin{eqnarray}
& &y^{'a}=y^{a} + \xi^{a}(x,y)
\hspace{1cm}  ({\rm O}(\xi^{2}) \ll 1), \nonumber\\
& &x^{' \mu}= x^{\mu},                \label{infff}
\end{eqnarray}
where $\xi^{a}$ is an arbitrary, infinitesimal function
of $x^{\mu}$ and $y^{a}$.
Thus we have
\begin{equation}
{\partial y^{'a}\over \partial y^{c}}
=\delta_{c}^{\ a}
+ {\partial \xi^{a}\over \partial y^{c}},
\end{equation}
or 
\begin{equation}
{\partial y^{c}\over \partial y^{'a}}
=\delta_{a}^{\ c}
- {\partial \xi^{c}\over \partial y^{a}}
 + \cdots,                      \label{idd}
\end{equation}
where $\cdots$ means terms of ${\rm O}(\xi^{2})$.
Let us expand the left hand side of the equation (\ref{coeff1})
using (\ref{infff}). It becomes
\begin{equation}
\phi'_{a b}(x',y+\xi)=\phi'_{a b}(x,y)
+ \xi^{c}{\partial \over \partial y^{c}}\phi_{a b}(x,y)
+ \cdots,                  \label{left}
\end{equation}
whereas the right hand side becomes
\begin{eqnarray}
\Big( {\partial y^{c}\over \partial y^{'a}}\Big)
\Big( {\partial y^{d}\over \partial y^{'b}}\Big)
\phi_{c d}(x,y)
&=&\Big( \delta_{a}^{\ c}
- {\partial \xi^{c}\over \partial y^{a}} + \cdots \Big)
\Big( \delta_{b}^{\ d}
- {\partial \xi^{d}\over \partial y^{b}} + \cdots \Big)
\phi_{c d}(x,y)            \nonumber\\
&=&\phi_{a b}(x,y)-
{\partial \xi^{c}\over \partial y^{a}}\phi_{c b}(x,y)
-{\partial \xi^{c}\over \partial y^{b}}\phi_{a c}(x,y)
+ \cdots.                      \label{heya}
\end{eqnarray}
From (\ref{left}) and (\ref{heya}), we therefore have
\begin{eqnarray}
\delta \phi_{a b}(x,y)
&\equiv &\phi'_{a b}(x,y)- \phi_{a b}(x,y) \nonumber\\
&=&-\xi^{c}{\partial \over \partial y^{c}} \phi_{a b}(x,y)
   -{\partial \xi^{c}\over \partial y^{a}}\phi_{c b}(x,y)
   -{\partial \xi^{c}\over \partial y^{b}}
    \phi_{a c}(x,y)                    \nonumber\\
&=&-[ \xi, \phi ]_{{\rm L} a b},           \label{finn}
\end{eqnarray}
where the subscript ${}_{\rm L}$ denotes the Lie derivative
of $\phi_{a b}$ along the vector field
$\xi=\xi^{c}\partial / \partial y^{a}$, i.e.
\begin{equation}
[ \xi, \phi ]_{{\rm L} a b}
=\xi^{c}{\partial \over \partial y^{c}} \phi_{a b}
   +{\partial \xi^{c}\over \partial y^{a}}\phi_{c b}
   +{\partial \xi^{c}\over \partial y^{b}} \phi_{a c}.
\end{equation}
It is a straightforward exercise to derive the infinitesimal 
transformation properties $A_{\mu}^{\ a}$ and $\gamma_{\mu\nu}$ 
from the finite transformations (\ref{coeff2}) and (\ref{coeff3}).
They are found to be
\begin{eqnarray}
\delta A_{\mu}^{\ a}(x,y)
&=&-{\partial \xi^{a}\over \partial x^{\mu}}
   + [A_{\mu},\ \xi ]_{{\rm L}}^{a} \nonumber\\ 
&=&-D_{\mu}\xi^{a},      \label{hi}\\
\delta \gamma_{\mu\nu}(x,y)
&=&-[ \xi, \gamma_{\mu\nu}]_{{\rm L}},        \label{hia}
\end{eqnarray} 
where the Lie brackets are defined as
\begin{eqnarray}
& &[A_{\mu},\ \xi ]_{{\rm L}}^{a}= A_{\mu}^{\ c}
    {\partial \xi^{a}\over \partial y^{c}}
    -\xi^{c}{\partial \over \partial y^{c}}A_{\mu}^{\ a}, \nonumber\\ 
& &[ \xi, \gamma_{\mu\nu}]_{{\rm L}}=
\xi^{a}{\partial \over \partial y^{a}}\gamma_{\mu\nu}.
\end{eqnarray} 
The equations (\ref{finn}), (\ref{hi}), and (\ref{hia})
show that $\phi_{ab}$, $A_{\mu} ^ { \ a}$, and $\gamma_{\mu\nu}$
behave like a tensor field, gauge fields, and scalar fields
under the diff$N_{2}$ transformation, respectively.
These are the equations (\ref{var2}),(\ref{var1}), and (\ref{var3})
in section \ref{s2}. 

Let us also derive the following transformation 
properties of $D_{\mu}\phi_{ab}$ and 
$F_{\mu\nu}^{\ \ a}$ under the diff$N_{2}$ transformations,
\begin{eqnarray}
& &\delta ( D_{\mu}\phi_{a b} )
=-[ \xi, D_{\mu}\phi]_{{\rm L} ab}, \label{fifx}\\    
& &\delta F_{\mu\nu}^{\ \ a}
=-[\xi, F_{\mu\nu}]_{\rm L}^{a},         \label{fify}
\end{eqnarray}
as given in equations (\ref{fifc}) and (\ref{fifa}) 
in section \ref{s2}.
Notice that 
\begin{eqnarray}
\delta ( D_{\mu}\phi_{a b} )
&=&\partial_\mu (\delta \phi_{a b}) 
-(\delta A_{\mu} ^ { \ c})(\partial_c \phi_{a b})
-A_{\mu} ^ { \ c}\partial_c (\delta \phi_{a b})
-\partial_a ( \delta A_{\mu} ^ { \ c} )\phi_{b c}\nonumber\\
& &-( \partial_a A_{\mu} ^ { \ c} )\delta \phi_{b c}
-\partial_b ( \delta A_{\mu} ^ { \ c} )\phi_{a c}
-( \partial_b A_{\mu} ^ { \ c} )\delta \phi_{a c} \nonumber\\
&=&-\partial_\mu \Big( [ \xi , \phi]_{{\rm L} ab} \Big)
+(D_{\mu}\xi^{c})( \partial_c \phi_{a b} )
+A_{\mu} ^ { \ c}\partial_c \Big( [ \xi , \phi ]_{{\rm L} ab} \Big)
+\partial_a ( D_{\mu}\xi^{c} ) \phi_{ bc} \nonumber\\
& &+(\partial_a A_{\mu} ^ { \ c})[ \xi , \phi ]_{{\rm L} bc}
+\partial_b ( D_{\mu}\xi^{c} ) \phi_{ ac}
+(\partial_b A_{\mu} ^ { \ c})[ \xi , \phi ]_{{\rm L} ac}\nonumber\\
&=&-\partial_\mu \Big( [ \xi , \phi]_{{\rm L} ab} \Big)
+[ A_{\mu} ,  [\xi, \phi ]_{\rm L}  ]_{{\rm L} ab}
+[D_{\mu}\xi, \phi ]_{{\rm L} ab},
\end{eqnarray}
using the equations (\ref{finn}) and (\ref{hi}),
and the following Lie brackets
\begin{eqnarray}
& &[ A_{\mu} ,  [\xi, \phi ]_{\rm L}   ]_{{\rm L} ab}
=A_{\mu} ^ { \ c}\partial_c \Big( [ \xi , \phi ]_{{\rm L} ab} \Big)
+(\partial_a A_{\mu} ^ { \ c})[ \xi , \phi ]_{{\rm L} bc}
+(\partial_b A_{\mu} ^ { \ c})[ \xi , \phi ]_{{\rm L} ac}, \\
& &[D_{\mu}\xi, \phi ]_{{\rm L} ab}
=(D_{\mu}\xi^{c})( \partial_c \phi_{a b} )
+\partial_a ( D_{\mu}\xi^{c} ) \phi_{ bc} 
+\partial_b ( D_{\mu}\xi^{c} ) \phi_{ ac}.
\end{eqnarray}
Using the Leibniz rule of the derivative $\partial_{\mu}$
\begin{equation}
\partial_\mu \Big( [ \xi , \phi]_{{\rm L} ab} \Big)
=[\partial_\mu  \xi , \phi]_{{\rm L} ab} 
+[ \xi , \partial_\mu \phi]_{{\rm L} ab},
\end{equation}
and the properties of the Lie bracket
\begin{eqnarray}
& &[ D_{\mu}\xi, \phi  ]_{{\rm L} ab}
= [\partial_\mu  \xi ,\phi ]_{{\rm L} ab} 
- [  [A_{\mu}, \xi ]_{\rm L},\phi  ]_{{\rm L} ab}, \\
& &[ A_{\mu} ,  [\xi, \phi ]_{\rm L}  ]_{{\rm L} ab}
=-[\xi, [ \phi , A_{\mu}]_{\rm L}  ]_{{\rm L} ab}
- [\phi , [A_{\mu}, \xi ]_{\rm L}  ]_{{\rm L} ab},
\end{eqnarray}
we find that
\begin{eqnarray}
\delta ( D_{\mu}\phi_{a b} )
&=&- [\xi, \partial_{\mu}\phi    ]_{{\rm L} ab} 
   + [ \xi, [ A_{\mu},   \phi  ]_{\rm L}  ]_{{\rm L} ab}  \nonumber\\
&=&- [  \xi, D_\mu \phi ]_{{\rm L} ab}.
\end{eqnarray}
The infinitesimal variation of $F_{\mu\nu}^{\ \ a}$ can be obtained
in a similar way. It becomes
\begin{eqnarray}
\delta F_{\mu\nu}^{\ \ a}
&=&\partial_\mu (\delta A_{\nu}^{\ a} )
- [ \delta A_{\mu}, A_{\nu}  ]_{\rm L}^{a}
-(\mu \leftrightarrow\nu )  \nonumber\\
&=&-\partial_\mu \Big\{   
\partial_\nu \xi^{a} - [ A_{\nu},  \xi ]_{\rm L}^{a}\Big\}
+[ D_{\mu}\xi, A_{\nu}  ]_{\rm L}^{a}
-(\mu \leftrightarrow\nu )  \nonumber\\
&=&\partial_\mu \Big( [ A_{\nu}, \xi ]_{\rm L}^{a}  \Big)
+[   D_{\mu}\xi, A_{\nu}  ]_{\rm L}^{a}
-(\mu \leftrightarrow\nu ).           \label{stop}
\end{eqnarray}
Using the following identities 
\begin{eqnarray}
& &\partial_\mu \Big( [ A_{\nu}, \xi ]_{\rm L}^{a}  \Big)
  =[ \partial_\mu A_{\nu}, \xi  ]_{\rm L}^{a}
  +[A_{\nu}, \partial_\mu \xi ]_{\rm L}^{a},  \\
& &[ D_{\mu}\xi, A_{\nu} ]_{\rm L}^{a}
 =-[ A_{\nu}, D_{\mu}\xi  ]_{\rm L}^{a}   
 =-[ A_{\nu}, \partial_\mu \xi  ]_{\rm L}^{a}
 +[ A_{\nu}, [  A_{\mu},  \xi ]_{\rm L}^{} ]_{\rm L}^{a},
\end{eqnarray}
we find that
\begin{eqnarray}
\delta F_{\mu\nu}^{\ \ a}
&=&[\partial_\mu A_{\nu}, \xi ]_{\rm L}^{a}
+[A_{\nu}, [A_{\mu}, \xi   ]_{\rm L}^{}   ]_{\rm L}^{a} 
-(\mu \leftrightarrow\nu )  \nonumber\\
&=&[\partial_\mu A_{\nu}-\partial_\nu A_{\mu} , \xi ]_{\rm L}^{a}
+[A_{\nu}, [A_{\mu}, \xi   ]_{\rm L}^{}   ]_{\rm L}^{a} 
-[A_{\mu}, [A_{\nu}, \xi   ]_{\rm L}^{}   ]_{\rm L}^{a} \nonumber\\
&=&-[ \xi , \partial_\mu A_{\nu}-\partial_\nu A_{\mu} 
   -[ A_{\mu}, A_{\nu} ]_{\rm L}^{}  ]_{\rm L}^{a}     \nonumber\\
&=&-[ \xi , F_{\mu\nu}]_{\rm L}^{a},
\end{eqnarray}
where we used the Jacobi identity
\begin{equation}
[ A_{\nu}, [ A_{\mu},  \xi ]_{\rm L}^{}  ]_{\rm L}^{a}
=-[ A_{\mu}, [\xi, A_{\nu}  ]_{\rm L}^{} ]_{\rm L}^{a}
-[  \xi, [ A_{\nu}, A_{\mu} ]_{\rm L}^{} ]_{\rm L}^{a}.
\end{equation}
Therefore it follows that
\begin{equation}
\delta F_{\mu\nu}^{\ \ a}=-[ \xi , F_{\mu\nu}]_{\rm L}^{a}.
\end{equation}

\section{The Einstein-Hilbert Action in the (2,2)-Splitting}
\label{a2}

In this Appendix we shall describe the procedure
of obtaining the scalar curvature (\ref{too}) 
in section \ref{s2} of 4-dimensional spacetime.
In the (2,2)-splitting of spacetime, we regard the 4-dimensional 
spacetime in question as a local product of a (1+1)-dimensional 
base manifold $M_{1+1}$ and a 2-dimensional fibre space $N_{2}$, 
whose basis vector fields are
$\partial_\mu$ and $\partial_a$, respectively.
The (1+1)-dimensional space orthogonal to the fibre space $N_{2}$
is called the horizontal space, generated by
the vector fields that can be written as linear combinations 
of $\partial_\mu$ and $\partial_a$,
\begin{equation}
\hat{\partial}_\mu = \partial_\mu
- A_{\mu} ^ {\ a}\partial_a,        \label{covz}
\end{equation}
where $A_{\mu} ^ {\ a}$ are the coefficient functions 
of $(x^{\mu},y^{a})$. Let $\hat{\partial}_a$ be such that
\begin{equation}
\hat{\partial}_a =\partial_a.     \label{cova}
\end{equation}
The basis ($\hat{\partial}_\mu, \hat{\partial}_a$) is called
the horizontal lift basis\cite{cho},
and the metric coefficients in this basis are given by
\begin{equation}
\hat{g}_{A B}=\left(\matrix{\gamma_{\mu\nu}  &
0 \cr 0 & \phi_{ab} \cr }\right).         \label{block}
\end{equation}
Here $\gamma_{\mu\nu}$ and $\phi_{ab}$ are the metrics
of the horizontal space and the fibre space, respectively,
and that 
\begin{equation}
\hat{g}_{\mu a}=0 
\end{equation}
is a consequence of the orthogonality of the horizontal space 
and the fibre space. 
In general, the horizontal lift basis vector fields are anholonomic, 
\begin{equation}
[\hat{\partial}_A, \hat{\partial}_B]=f_{A B} ^ { \
\  \ C}\hat{\partial}_C,
\end{equation}
where $f_{A B} ^ { \  \  \ C}$ are the structure functions
satisfying the relation
\begin{equation}
f_{A B} ^ { \  \  \ C}=f_{[A B]} ^ { \ \  \  \ C}.
\end{equation}
These structure functions are found to be 
\begin{eqnarray}
& &f_{\mu\nu} ^ { \  \ a}=-f_{\nu\mu} ^ { \  \ a}
  =-F_{\mu\nu} ^ { \  \ a},            \nonumber\\
& &f_{\mu a} ^ { \  \ b}=-f_{a \mu}^{ \  \ b}
  =\partial_a A_{\mu} ^ { \ b}, \label{struca}
\end{eqnarray}
and zero otherwise.
Here $F_{\mu\nu} ^ { \  \ a}$ is defined as
\begin{equation}
F_{\mu\nu} ^ { \  \ a}
=\partial_\mu A_{\nu} ^ { \ a}-\partial_\nu
A_{\mu} ^ { \ a} - [A_{\mu}, A_{\nu}]_{\rm L}^{a},  \nonumber\\
\end{equation}
as before.

The formulae of the connection coefficients and
the curvature tensors in the anholonomic basis
can be found in \cite{cho,mtw}. In the horizontal lift basis
where the metric is given by (\ref{block}), they are given by 
\begin{eqnarray}
& &\hat{\Gamma}_{A B}^{ \ \ \ C}={1\over 2}\hat{g}^{C D}
  \Big( \hat{\partial}_{A}\hat{g}_{B D}
  +\hat{\partial}_{B}\hat{g}_{A D}
  - \hat{\partial}_{D}\hat{g}_{A B} \Big)
  +{1\over 2}\hat{g}^{C D}\Big( f_{A B D} - f_{B D A}
 - f_{A D B} \Big),                        \label{con1}\\
& &\hat{R}_{A B C}^{ \ \ \ \ \ D}
  = \hat{\partial}_{A}^{}\hat{\Gamma}_{B C}^{\ \ \ D}
   -\hat{\partial}_{B}^{}\hat{\Gamma}_{A C}^{\ \ \ D}
   + \hat{\Gamma}_{A E}^{\ \ \ D}\hat{\Gamma}_{B C}^{\ \ \ E}
   - \hat{\Gamma}_{B E}^{\ \ \ D}\hat{\Gamma}_{A C}^{\ \ \ E}
   -f_{A B}^{\ \ \ E}\hat{\Gamma}_{E C}^{\ \ \ D},  \label{con2}\\
& &\hat{R}_{A C}=\hat{R}_{A B C}^{\ \ \ \ \ B}, \label{con3}\\
& &R=\hat{g}^{A C}\hat{R}_{A C},      \label{con4}
\end{eqnarray}
where $f_{A B C}=\hat{g}_{C D}f_{A B}^{\ \ \ D}$.
Notice that in this basis
$\hat{\Gamma}_{A B}^{ \ \ \ C}$ is, in general, not symmetric:
\begin{equation}
\hat{\Gamma}_{A B}^{ \ \ \ C}\neq \hat{\Gamma}_{B A}^{ \ \ \ C}
\ \ \ \ {\rm if}\ \ \ \ f_{A B}^{\ \ \ C}\neq 0.
\end{equation}
In components the connection coefficients are given by
\begin{eqnarray}
& &\hat{\Gamma}_{\mu\nu}^{\ \ \alpha}
   ={1\over 2}\gamma^{\alpha\beta}\Big(
   \hat{\partial}_{\mu}\gamma_{\nu\beta}
   + \hat{\partial}_{\nu}\gamma_{\mu\beta}
   -\hat{\partial}_{\beta}\gamma_{\mu\nu}  \Big), \label{com1}\\
& &\hat{\Gamma}_{\mu\nu}^{\ \ a}
   = -{1\over 2}\phi^{a b}\partial_{b}\gamma_{\mu\nu}
  - {1\over 2}F_{\mu\nu}^{\ \ a},                \label{com2}\\
& &\hat{\Gamma}_{\mu a}^{\ \ \nu}=\hat{\Gamma}_{ a\mu}^{\ \ \nu}
   ={1\over 2}\gamma^{\nu\alpha}\partial_{a}\gamma_{\mu\alpha}
   + {1\over 2}\gamma^{\nu\alpha}
    \phi_{a b}F_{\mu\alpha}^{\ \ b}, \         \label{com3}\\
& &\hat{\Gamma}_{\mu a}^{\ \ b}
   ={1\over 2}\phi^{b c}\hat{\partial}_{\mu}\phi_{a c}
   +{1\over 2}\partial_{a}A_{\mu}^{\ b}
   - {1\over 2}\phi^{b c}
    \phi_{a e}\partial_{c}A_{\mu}^{\ e} \nonumber\\
& &\hspace{.8cm}={1\over 2}\phi^{b c}D_{\mu}\phi_{a c}
    +\partial_{a}A_{\mu}^{\ b},     \label{com4}\\
& &\hat{\Gamma}_{a\mu}^{\ \ b}
   ={1\over 2}\phi^{b c}\hat{\partial}_{\mu}\phi_{a c}
   -{1\over 2}\partial_{a}A_{\mu}^{\ b}
   -{1\over 2}\phi^{b c}\phi_{a e}
   \partial_{c}A_{\mu}^{\ e}  \nonumber\\
& &\hspace{.8cm}={1\over 2}\phi^{b c}D_{\mu}\phi_{a c}, \label{com5}\\
& &\hat{\Gamma}_{a b}^{\ \ \mu}
  =-{1\over 2}\gamma^{\mu\nu}\hat{\partial}_{\nu}\phi_{a b}
  +{1\over 2}\gamma^{\mu\nu}\phi_{a c}\partial_{b}A_{\nu}^{\ c}
  +{1\over 2}\gamma^{\mu\nu}\phi_{b c}
   \partial_{a}A_{\nu}^{\ c}   \nonumber\\
& &\hspace{.8cm}=-{1\over 2}
   \gamma^{\mu\nu}D_{\nu}\phi_{a b},  \label{com6}\\
& &\hat{\Gamma}_{a b}^{\ \ c}={1\over 2}\phi^{c d}\Big(
  \partial_{a}\phi_{b d} + \partial_{b}\phi_{a d}
  - \partial_{d}\phi_{a b}\Big),             \label{com7}
\end{eqnarray}
where $\hat{\partial}_{\mu}\gamma_{\alpha\beta}$ and
$\hat{\partial}_{\mu}\phi_{a b}$ are given by
\begin{eqnarray}
& &\hat{\partial}_{\mu}\gamma_{\alpha\beta}
=\partial_{\mu}\gamma_{\alpha\beta}
- A_{\mu}^{\ c}\partial_{c}\gamma_{\alpha\beta},   \label{hata}\\
& &\hat{\partial}_{\mu}\phi_{a b}
=\partial_{\mu}\phi_{a b}
-A_{\mu}^{\ c}\partial_{c}\phi_{a b},              \label{hatb}
\end{eqnarray}
and $D_{\mu}\phi_{a b}$ is defined as
\begin{eqnarray}
D_{\mu}\phi_{a b}&=&\partial_{\mu}\phi_{a b}
-A_{\mu}^{\ c}\partial_{c}\phi_{a b}
-(\partial_{a}A_{\mu}^{\ c})\phi_{b c}
-(\partial_{b}A_{\mu}^{\ c})\phi_{a c}  \nonumber\\
&=&\partial_{\mu}\phi_{a b} - [A_{\mu}, \phi ]_{{\rm L} a b}.
\end{eqnarray}
Here $[A_{\mu}, \phi ]_{{\rm L} a b}$ is the Lie derivative of
$\phi_{a b}$ along the vector field
$A_{\mu}=A_{\mu}^{\ a}\partial_{a}$.
From (\ref{com1}), $\cdots$, (\ref{com7}),
we also find the following useful identities,
\begin{eqnarray}
& &\hat{\Gamma}_{\mu\nu}^{\ \ \mu}=\hat{\Gamma}_{\nu\mu}^{\ \ \mu}
  ={1\over 2}\gamma^{\alpha\beta}
   \hat{\partial}_{\nu}\gamma_{\alpha\beta}, \label{contra}\\
& &\hat{\Gamma}_{\mu a}^{\ \ \mu}=\hat{\Gamma}_{a \mu}^{\ \ \mu}
  ={1\over 2}\gamma^{\alpha\beta}
  \partial_{a}\gamma_{\alpha\beta},          \label{contrab}\\
& &\hat{\Gamma}_{\nu a}^{\ \ a}
  ={1\over 2}\phi^{a b}D_{\nu}\phi_{a b}
   +\partial_{a}A_{\nu}^{\ a},               \label{contrac}\\
& &\hat{\Gamma}_{a \nu}^{\ \ a}
   ={1\over 2}\phi^{a b}D_{\nu}\phi_{a b},      \label{contrad}\\
& &\hat{\Gamma}_{a b}^{\ \ a}
={1\over 2}\phi^{a c}\partial_{b}\phi_{a c}.        \label{use}
\end{eqnarray}
The scalar curvature $R$ of the 4-dimensional spacetime is given by
\begin{eqnarray}
R&=&\hat{g}^{A B}\hat{R}_{AB}    \nonumber\\
 &=&\gamma^{\mu\nu}\hat{R}_{\mu\nu}
    +\phi^{a c}\hat{R}_{a c},           \label{forma}
\end{eqnarray}
where $\hat{R}_{\mu\nu}$ and $\hat{R}_{a c}$ are given by
\begin{eqnarray}
& &\hat{R}_{\mu\nu}=\hat{R}_{\mu \alpha \nu}^{\ \ \ \ \alpha}
+\hat{R}_{\mu a \nu}^{\ \ \ \ a},            \nonumber\\
& &\hat{R}_{a c}=\hat{R}_{a b c}^{\ \ \ \ b}
+\hat{R}_{a \alpha c}^{\ \ \ \ \alpha},        \label{formb}
\end{eqnarray}
respectively. 
The quantities $\gamma^{\mu\nu}\hat{R}_{\mu\nu}$ and 
$\phi^{a c}\hat{R}_{a c}$ are found to be
\begin{eqnarray}
\gamma^{\mu\nu}\hat{R}_{\mu\nu}&=&\gamma^{\mu\nu}\tilde{R}_{\mu\nu}
+{1\over 2}\gamma^{\mu\nu}\gamma^{\alpha\beta}\phi_{a b}
  F_{\mu\alpha} ^ { \  \ a} F_{\nu\beta}^{\ \ b}
+{1\over 4}\gamma^{\mu\nu}\phi^{a b}\phi^{c d}
\Big\{
 (D_{\mu}\phi_{a c})(D_{\nu}\phi_{b d})  \nonumber\\
& &-(D_{\mu}\phi_{a b})(D_{\nu}\phi_{c d}) \Big\}
+{1\over 2}\Big( \hat{\nabla}_{\mu}
   +\hat{\Gamma}_{c \mu}^{\ \ c}\Big) j^{\mu}
   +{1\over 2}\Big( \hat{\nabla}_{a}
   +\hat{\Gamma}_{\alpha a}^{\ \ \alpha}\Big) j^{a},  \label{qqq}\\
\phi^{a c}\hat{R}_{a c}&=&\phi^{a c}R_{a c}
-{1\over 4}\gamma^{\mu\nu}\gamma^{\alpha\beta}
  \phi_{a b}F_{\mu\alpha} ^ { \  \ a}
  F_{\nu\beta}^{\ \ b}
+{1\over 4}\phi ^ {a b}\gamma^{\mu\nu}\gamma^{\alpha\beta}
\Big\{
  (\partial_{a}\gamma_{\mu \alpha})
  (\partial_{b}\gamma_{\nu\beta})              \nonumber\\
& &-(\partial_{a}\gamma_{\mu\nu})
  (\partial_{b}\gamma_{\alpha\beta})    \Big\}
  +{1\over 2}\Big( \hat{\nabla}_{\mu}
  +\hat{\Gamma}_{c \mu}^{\ \ c}\Big) j^{\mu}
 +{1\over 2}\Big( \hat{\nabla}_{a}
 +\hat{\Gamma}_{\alpha a}^{\ \ \alpha}\Big) j^{a}, \label{rrr}
\end{eqnarray}
respectively, and the scalar curvature $R$ is given by their sum,
\begin{eqnarray}
R&=&\gamma^{\mu\nu}\tilde{R}_{\mu\nu} + \phi^{a c}R_{a c}
 + {1\over 4}\gamma^{\mu\nu}\gamma^{\alpha\beta}
  \phi_{a b}F_{\mu\alpha} ^ { \  \ a}
  F_{\nu\beta}^{\ \ b}
+{1\over 4}\gamma^{\mu\nu}\phi ^ {a b}\phi ^ {c d}\Big\{
  (D_{\mu}\phi_{a c})(D_{\nu}\phi_{b d})       \nonumber\\
& & -(D_{\mu}\phi_{a b})(D_{\nu}\phi_{c d}) \Big\}
  +{1\over 4}\phi ^ {a b}\gamma^{\mu\nu}
  \gamma^{\alpha\beta}\Big\{
 (\partial_{a}\gamma_{\mu \alpha})(\partial_{b}\gamma_{\nu\beta})
 -(\partial_{a}\gamma_{\mu \nu})(\partial_{b}
  \gamma_{\alpha\beta}) \Big\}             \nonumber\\
& & +(\hat{\nabla}_{\mu}+\hat{\Gamma}_{c \mu}^{\ \ c})j^{\mu}
 +(\hat{\nabla}_{a}
 +\hat{\Gamma}_{\alpha a}^{\ \ \alpha})j^{a}.  \label{to}
\end{eqnarray}
We summarize the notations below:\\
1. $\tilde{R}_{\mu\nu}$ is defined as
\begin{equation}
\tilde{R}_{\mu\nu}=
  \hat{\partial}_{\mu}^{}\hat{\Gamma}_{\alpha \nu}^{\ \ \alpha}
 -\hat{\partial}_{\alpha}^{}
  \hat{\Gamma}_{\mu \nu}^{\ \ \alpha}
 +\hat{\Gamma}_{\mu \beta}^{\ \ \alpha}
  \hat{\Gamma}_{\alpha \nu}^{\ \ \beta}
 -\hat{\Gamma}_{\beta \alpha}^{\ \ \beta}
  \hat{\Gamma}_{\mu \nu}^{\ \ \alpha}.          \label{gric}
\end{equation}
Notice that the connection coefficients 
$\hat{\Gamma}_{\mu \nu}^{\ \ \alpha}$ are given by
\begin{equation}
\hat{\Gamma}_{\mu \nu}^{\ \ \alpha}
={1\over 2}\gamma^{\alpha\beta}\Big(
 \hat{\partial}_{\mu}\gamma_{\nu\beta}
 + \hat{\partial}_{\nu}\gamma_{\mu\beta}
 -\hat{\partial}_{\beta}\gamma_{\mu\nu}  \Big),   \label{ok}
\end{equation}
and that $\tilde{R}_{\mu\nu}$ is defined using the metric
$\gamma_{\mu\nu}$ of the horizontal space and the horizontal
vector fields $\hat{\partial}_{\mu}$ only.
Therefore $\tilde{R}_{\mu\nu}$ is the Ricci tensor of 
the horizontal space. \\ 
\noindent
2. $R_{a c}$ is defined as
\begin{equation}
R_{a c}=\partial_{a}^{}\hat{\Gamma}_{b c}^{\ \ b}
-\partial_{b}^{}\hat{\Gamma}_{a c}^{\ \ b}
+\hat{\Gamma}_{a d}^{\ \ b}
 \hat{\Gamma}_{b c}^{\ \ d}
-\hat{\Gamma}_{d b}^{\ \ d}\hat{\Gamma}_{a c}^{\ \ b}. \label{qten}
\end{equation}
The connection coefficients $\hat{\Gamma}_{a b}^{\ \ c}$ 
of the fibre space $N_{2}$ are given by
\begin{equation}
\hat{\Gamma}_{a b}^{\ \ c}
  ={1\over 2}\phi^{c d}\Big(
  \partial_{a}\phi_{b d} + \partial_{b}\phi_{a d}
 -\partial_{d}\phi_{a b}\Big).
\end{equation}
Therefore $R_{a c}$ is the Ricci tensor of $N_{2}$.\\ 
\noindent
3. For completeness, we quote here the definition of the 
diff$N_{2}$-field strength $F_{\mu\nu} ^ { \  \ a}$ and 
the diff$N_{2}$-covariant derivative $D_{\mu}\phi_{a b}$,
\begin{eqnarray}
F_{\mu\nu} ^ { \  \ a}
&=&\partial_\mu A_{\nu} ^ { \ a}-\partial_\nu
A_{\mu} ^ { \ a} - [A_{\mu}, A_{\nu}]_{\rm L}^{a}, \nonumber\\
D_{\mu}\phi_{a b}
&=&\partial_{\mu}\phi_{a b}
   - [A_{\mu}, \phi ]_{{\rm L} a b}.       
\end{eqnarray}
As is shown in Appendix \ref{a1}, $F_{\mu\nu} ^ { \  \ a}$ and 
$D_{\mu}\phi_{a b}$
transform covariantly under the diff$N_{2}$ transformations.\\ 
\noindent
4.The derivative operators $\hat{\nabla}_{\mu}$
and $\hat{\nabla}_{a}$ are the natural derivative operators
compatible with $\gamma_{\alpha\beta}$ and $\phi_{b c}$.
In the horizontal lift basis, the compatibility conditions can 
be written as
\begin{eqnarray}
& &\hat{\nabla}_{\mu}\gamma_{\alpha\beta}
=\hat{\partial}_{\mu}\gamma_{\alpha\beta}
-\hat{\Gamma}_{\mu\alpha}^{\ \ \delta}\gamma_{\delta\beta}
-\hat{\Gamma}_{\mu\beta}^{\ \ \delta}
\gamma_{\alpha\delta}=0,                \label{compati}\\
& &\hat{\nabla}_{a}\phi_{b c}
=\partial_{a}\phi_{b c}
-\hat{\Gamma}_{a b}^{\ \ d}\phi_{d c}
-\hat{\Gamma}_{a c}^{\ \ d}\phi_{b d}=0,     \label{compatia}
\end{eqnarray}
respectively. These properties can be also checked using
the formulae (\ref{com1}) and (\ref{com7}). \\ \noindent
5.We also defined $j^{\mu}$ and $j^{a}$ as
\begin{eqnarray}
& &j^{\mu}=\gamma^{\mu\nu}
        \phi^{a b}D_{\nu}\phi_{a b},           \label{jeimu}\\
& &j^{a}=\phi^{a b} \gamma^{\mu\nu}
        \partial_{b}\gamma_{\mu\nu}.           \label{jeiei}
\end{eqnarray}           
Thus $\hat{\nabla}_{\mu}j^{\mu}$ and $\hat{\nabla}_{a}j^{a}$ 
are given by
\begin{eqnarray}
& &\hat{\nabla}_{\mu}j^{\mu}
=\hat{\partial}_{\mu}j^{\mu}
+\hat{\Gamma}_{\mu\nu}^{\ \ \mu}j^{\nu},     \label{derivz}\\
& &\hat{\nabla}_{a}j^{a}
=\partial_{a}j^{a}
+\hat{\Gamma}_{a b}^{\ \ a}j^{b},       \label{deriva}
\end{eqnarray}
respectively. 

The last two terms containing $j^{\mu}$ and $j^{a}$ 
in (\ref{to}) turn out to be surface terms in the action integral, 
as can be seen in what follows. Let us consider the
term that contains $j^{\mu}$. It becomes 
\begin{eqnarray}
\sqrt{-\gamma}\sqrt{\phi} \,
\Big(\hat{\nabla}_{\mu}+\hat{\Gamma}_{c \mu}^{\ \ c}\Big)j^{\mu}
&=&\sqrt{-\gamma}\sqrt{\phi} \, \Big(
   \hat{\partial}_{\mu} j^{\mu}
   +\hat{\Gamma}_{\nu\mu}^{\ \ \nu}j^{\mu}
   +\hat{\Gamma}_{c \mu}^{\ \ c}j^{\mu} \Big) \nonumber\\
&=&\sqrt{-\gamma}\sqrt{\phi} \, \Big(
\partial_{\mu}j^{\mu}- A_{\mu}^{\ a}\partial_{a}j^{\mu}
+\hat{\Gamma}_{\nu\mu}^{\ \ \nu}j^{\mu}
+\hat{\Gamma}_{c \mu}^{\ \ c}j^{\mu} \Big). \label{firsta}
\end{eqnarray}
The first term in the right hand side of 
(\ref{firsta}) can be written as
\begin{equation}
\sqrt{-\gamma}\sqrt{\phi} \, \partial_{\mu}j^{\mu}
=-{1\over 2}\sqrt{-\gamma}\sqrt{\phi} \,
 \Big( \gamma^{\alpha\beta}\partial_{\mu}\gamma_{\alpha\beta}
 +\phi^{a b}\partial_{\mu}\phi_{a b}\Big) j^{\mu}
 +\partial_{\mu}\Big(
  \sqrt{-\gamma}\sqrt{\phi} \, j^{\mu}\Big),  \label{firstb}
\end{equation}
and the second term becomes
\begin{eqnarray}
-\sqrt{-\gamma}\sqrt{\phi} \,
     A_{\mu}^{\ a}\partial_{a}j^{\mu}
&=&{1\over 2}\sqrt{-\gamma}\sqrt{\phi} \, A_{\mu}^{\ a} \Big(
 \gamma^{\alpha\beta}\partial_{a}\gamma_{\alpha\beta}
 +\phi^{b c}\partial_{a}\phi_{b c} \Big) j^{\mu}
+\sqrt{-\gamma}\sqrt{\phi} \,
( \partial_{a}A_{\mu}^{\ a} ) j^{\mu}       \nonumber\\
& &-\partial_{a} \Big( \sqrt{-\gamma}\sqrt{\phi} \,
  A_{\mu}^{\ a}j^{\mu}\Big).       \label{firstc}
\end{eqnarray}
Using the equations (\ref{contra}) and (\ref{contrad}), 
the last two terms in (\ref{firsta}) can be written as
\begin{eqnarray}
& &\sqrt{-\gamma}\sqrt{\phi} \, \Big(
\hat{\Gamma}_{\nu\mu}^{\ \ \nu}j^{\mu}
+\hat{\Gamma}_{c \mu}^{\ \ c}j^{\mu}  \Big)   \nonumber\\
&=&\sqrt{-\gamma}\sqrt{\phi} \,  \Big(
{1\over 2}\gamma^{\alpha\beta}\hat{\partial}_{\mu}
        \gamma_{\alpha\beta}
+{1\over 2}\phi^{a b}D_{\mu}\phi_{a b}\Big) j^{\mu} \nonumber\\
&=&\sqrt{-\gamma}\sqrt{\phi} \,  \Big(
{1\over 2}\gamma^{\alpha\beta}\hat{\partial}_{\mu}
        \gamma_{\alpha\beta}
+{1\over 2}\phi^{a b}\hat{\partial}_{\mu}\phi_{a b}
- \partial_{c}A_{\mu}^{\ c} \Big) j^{\mu}  \nonumber\\
&=&\sqrt{-\gamma}\sqrt{\phi} \, \Big(
{1\over 2}\gamma^{\alpha\beta}\partial_{\mu}
        \gamma_{\alpha\beta}
-{1\over 2}A_{\mu}^{\ c}
\gamma^{\alpha\beta}\partial_{c} \gamma_{\alpha\beta}
+{1\over 2}\phi^{a b}\partial_{\mu}\phi_{a b}
-{1\over 2}A_{\mu}^{\ c}
 \phi^{a b}\partial_{c} \phi_{a b}     
-\partial_{c}A_{\mu}^{\ c}\Big) j^{\mu}.   \label{firstd}
\end{eqnarray}
From these, we find that (\ref{firsta}) is indeed 
a total divergence,
\begin{equation}
\sqrt{-\gamma}\sqrt{\phi} \,
\Big(\hat{\nabla}_{\mu}+\hat{\Gamma}_{c \mu}^{\ \ c}\Big)j^{\mu}
=\partial_{\mu}\Big(
  \sqrt{-\gamma}\sqrt{\phi} \, j^{\mu}\Big)
  -\partial_{a} \Big( \sqrt{-\gamma}\sqrt{\phi} \,
  A_{\mu}^{\ a}j^{\mu}\Big).           \label{firste}
\end{equation}
Similarly, we can show that the following is also a total divergence,
\begin{equation}
\sqrt{-\gamma}\sqrt{\phi} \,
(\hat{\nabla}_{a}+\hat{\Gamma}_{\alpha a}^{\ \ \alpha})j^{a}
 =\partial_{a}\Big(
  \sqrt{-\gamma}\sqrt{\phi} \,  j^{a}\Big).   \label{diva}
\end{equation}
Thus the Einstein-Hilbert action integral in this (2,2)-KK 
formalism can be written as
\begin{eqnarray}
S&=&\int \! \! d^{2}x d^{2}y \,
\sqrt{-\gamma}\sqrt{\phi} \, \Big[
\gamma^{\mu\nu}\tilde{R}_{\mu\nu} + \phi^{a c}R_{a c}
 + {1\over 4}\gamma^{\mu\nu}\gamma^{\alpha\beta}
  \phi_{a b}F_{\mu\alpha} ^ { \  \ a}
  F_{\nu\beta}^{\ \ b}          \nonumber\\
& & +{1\over 4}\gamma^{\mu\nu}\phi ^ {a b}\phi ^ {c d}\Big\{
  (D_{\mu}\phi_{a c})(D_{\nu}\phi_{b d})
  -(D_{\mu}\phi_{a b})(D_{\nu}\phi_{c d}) \Big\}  \nonumber\\
& &  +{1\over 4}\phi ^ {a b}\gamma^{\mu\nu}
  \gamma^{\alpha\beta}\Big\{
 (\partial_{a}\gamma_{\mu \alpha})(\partial_{b}\gamma_{\nu\beta})
 -(\partial_{a}\gamma_{\mu \nu})(\partial_{b}
  \gamma_{\alpha\beta}) \Big\} \Big]  \nonumber\\
& &+\int \! \! d^{2}x d^{2}y \,  \Big[ 
\partial_{\mu}\Big(
  \sqrt{-\gamma}\sqrt{\phi} \, j^{\mu}\Big)
  -\partial_{a} \Big( \sqrt{-\gamma}\sqrt{\phi} \,
  A_{\mu}^{\ a}j^{\mu}\Big)
+\partial_{a}\Big(
  \sqrt{-\gamma}\sqrt{\phi} \,  j^{a}\Big) \Big].
\end{eqnarray}

\section{The Positive Definiteness of $\kappa^{2}$}
\label{a3}

In this section we will show that $\kappa^{2}$, 
defined in (\ref{kappa}) in section \ref{s4},
is positive definite. To this end, it is convenient to
work with the superspace indices $A'=(a c)$, $B'=(b d)$ such that
\setcounter{equation}{0}
\begin{equation}
\rho_{A'}=\rho_{a c}, \hspace{1cm}
\rho_{B'}=\rho_{b d}.
\end{equation}
Define the metric $G_{A' B'}$ on the superspace
and its inverse metric $G^{A' B'}$ by
\begin{eqnarray}
& &G_{A' B'}={1\over 2}(\rho_{a b}\rho_{c d}+
        \rho_{a d}\rho_{c b}),       \label{super}\\
& &G^{A' B'}={1\over 2}(\rho^{a b}\rho^{c d}
  +\rho^{a d}\rho^{c b}),            \label{supera}
\end{eqnarray}
respectively, so that the following relation
\begin{equation}
G^{A' E'}G_{E' B'}=\delta^{A'}_{\ B'}
\end{equation}
holds. Here the superspace Kronecker delta is defined as
\begin{equation}
\delta^{A'}_{\ B'}={1\over 2}(\delta^{a}_{\ b}\delta^{c}_{\ d}
 + \delta^{a}_{\ d}\delta^{c}_{\ b}).
\end{equation}
This supermetric raises and lowers the superindices
\begin{eqnarray}
& &G^{A' B'}\rho_{B'}=\rho^{A'}, \nonumber\\
& &G_{A' B'}\rho^{B'}=\rho_{A'}.
\end{eqnarray}
Then $\kappa^{2}$ can be written as 
\begin{eqnarray}
\kappa^{2}&=&{1\over 8}\rho^{a b}\rho^{c d}
(D_{-}\rho_{a c}) (D_{-}\rho_{b d})       \nonumber\\
&=&{1\over 8}G^{A' B'}(D_{-}\rho_{A'})(D_{-}\rho_{B'}).
\end{eqnarray}
The positive definiteness of $\kappa^{2}$
comes from the positive definiteness of the supermetric $G^{A' B'}$. 
Let us notice that, when $\rho^{a b}=\delta^{a b}$, this supermetric
becomes
\begin{equation}
G^{A' B'}={\rm diag}(+1, +1/  2, +1),
\end{equation}
which is positive definite.
By continuity, it follows that
\begin{equation}
\kappa^{2}\geq 0.
\end{equation}


\begin{thebibliography}{99}
\bibitem{park}{Q.H. Park, Phys. Lett. B {\bf 238}, 287 (1990);
Int. J. Mod. Phys. A {\bf 7}, 1415 (1992).}
\bibitem{yoon}{Y.M. Cho,  Q.H. Park, K.S. Soh, and J.H. Yoon,
  Phys. Lett. B {\bf 286}, 251 (1992).}
\bibitem{yoona}{J.H. Yoon, Phys. Lett. B {\bf 308}, 240 (1993).}
\bibitem{yoonb}{J.H. Yoon,{\it (1+1)-Dimensional Methods for General
  Relativity},
  in {\it Directions in General Relativity}, in
  Proceedings of the 1993 International Symposium, Vol. 2,
  ed. B.L. Hu and T.A. Jacobson (Cambridge University Press, 1993).}
\bibitem{solo}{J.H. Yoon, {\it in preparation}.}
\bibitem{bondi}{H. Bondi, M.G.J. van der Burg, and A.W.K. Metzner,
  Proc. R. Soc. Lond. A {\bf 269}, 21 (1962).}
\bibitem{sach}{R.K. Sachs, Proc. R. Soc. Lond.
  A {\bf 270}, 103 (1962); J. Math. Phys.
  {\bf 3}, 908 (1962);
  {\it Gravitational Radiation}, in {\it
  Relativity, Groups and Topology}, The 1963 Les Houches Lectures,
  ed. B. deWitt and C. deWitt (Gordon and Beach, New York, 1964).}
\bibitem{asht}{A. Ashtekar, {\it Asymptotic Quantization},
{\it Monographs and Textbooks in Physical Sciences}
(Bibliopolis, 1986)}.
\bibitem{ashta}{A. Ashtekar, J. Bi\v{c}ak, and B. Schmidt, 
{\it Asymptotic Structure of Symmetry 
Reduced General Relativity}, gr-qc/9608042}.
\bibitem{carlos}{A.E. Dominguez, C.N. Kozameh, and M. Ludvigsen, 
{\it Superselection Sectors in Asymptotic 
  Quantization of Gravity}, gr-qc/9609071}.
\bibitem{helmut}{H. Friedrich and J.M. Stewart,
  Proc. R. Soc. Lond. A {\bf 385}, 345 (1983).}
\bibitem{gold}{J.N. Goldberg, Found. Phys.
  {\bf 14}, 1211 (1984); {\it ibid.} {\bf 15}, 439 (1985); 
  {\it ibid.} {\bf 15}, 1075 (1985).}
\bibitem{torr}{C.G. Torre, Class. Quantum. Grav.
  {\bf 3}, 773 (1986).}
\bibitem{cho}{Y.M. Cho, J. Math. Phys.
  {\bf 16}, 2029 (1975).}
\bibitem{mtw}{C.W. Misner, K.S. Thorne, and J.A. Wheeler,
  {\it Gravitation}  (Freeman, 1972). }
\bibitem{geroch}{R. Geroch, in
{\it Asymptotic Structure of Space-time}, ed. P. Esposito
and L. Witten (Plenum, 1977).}
\bibitem{diff}{P.R. Garabedian,
{\it Partial Differential Equations} (John Wiley and Sons, 1966).}
\bibitem{unti}{E.T. Newman and T.W.J. Unti, J. Math. Phys.
  {\bf 3}, 891 (1962).}          
\bibitem{inv}{R.A. d'Inverno and J. Stachel, J. Math. Phys.
  {\bf 19}, 2447 (1978).}    
\bibitem{inva}{R.A. d'Inverno and J. Smallwood, Phys. Rev.
  D {\bf 22}, 1233 (1980).}
\bibitem{small}{J. Smallwood, J. Math. Phys.
  {\bf 24}, 599 (1983).}
\bibitem{newman+tod}{E.T. Newman and K.P Tod,
{\it Asymptotically Flat Space-Times}, in 
{\it General Relativity and Gravitation}, Vol.2,
ed. A. Held (Plenum Press, 1980)}.
\bibitem{brill}{D. Brill, Ann. Phys.
 {\bf 7}, 466 (1959); Nuovo Cimento Supp.
  {\bf 2}, 1 (1964).}
\bibitem{sssol}{J.H. Yoon, C.M. Kim, and S.K. Oh,
{\it A New Way of Looking at Exact Solutions:
Schwartzschild Solution in the (2,2) Kaluza-Klein Formalism 
of General Relativity}, {\it in preparation}. }
\bibitem{exact}{D. Kramer {\it et al}, 
{\it Exact Solutions of Einstein's Field Equations},
ed. E. Schmutzer (Cambridge University Press, 1980).}
\end{thebibliography}
\end{document}